\documentclass[11pt]{article}

\usepackage{amsmath,amsthm,latexsym,amssymb,amsfonts,epsfig}
\usepackage{cite}

\usepackage[utf8]{inputenc}

\usepackage{lscape}

\usepackage{graphicx} 
\usepackage{physics}
\usepackage{amssymb}
\usepackage{color}

\usepackage[hidelinks]{hyperref}
\usepackage{booktabs}

\usepackage{bm}

\numberwithin{equation}{section}
\allowdisplaybreaks[1] 

\title{{\sf Quantum Field Theory of}\\
{\sf Black Hole Perturbations with Backreaction}\\
{\sf II. Spherically symmetric 2nd order Einstein sector}}

\author{
{\sf J. Neuser}$^1$\thanks{{\sf jonas.neuser@fau.de}},
{\sf T. Thiemann}$^1$\thanks{{\sf
thomas.thiemann@gravity.fau.de}}\\
\\
{\sf $^1$ Institute for Quantum Gravity, FAU Erlangen -- N\"urnberg,}\\
{\sf Staudtstr. 7, 91058 Erlangen, Germany}\\
}
\date{{\small\sf \today}}

\begin{document}

\maketitle

\begin{abstract}
    In this second paper of our series we focus on the classical pure gravity sector of spherically 
    symmetric black hole perturbations and expand the reduced Hamiltonian to second order.
    To compare our manifestly gauge invariant formalism with established results in the literature
    we have to translate our results derived in Gullstrand-Painlev\'e gauge to the gauges used 
    in those works. After several canonical transformations we expectedly find exact agreement 
    with the Hamiltonian given by Moncrief which generates the Regge-Wheeler and Zerilli equations 
    of motion for the linear axial (often denoted odd) and polar (often denoted even) perturbations 
    respectively. This confirms the validity of our method which immediately 
    generalises to higher orders.
\end{abstract}

\section{Introduction}
In \cite{I} we developed a framework for the quantum field theory (QFT) of black hole perturbations 
that allows for quantum geometry backreaction. Therefore our notion of backreaction differs 
from the more common semiclassical notion
in which one treats matter quantum fields on a geometry background that has to self-consistently 
solve the semiclassical Einstein equations which result from replacing 
the classical matter energy momentum tensor by a suitable expectation value of its matter quantised
version. Thus this framework applies to a perturbative regime of 
quantum gravity proper and goes beyond QFT in curved spacetimes. 
The new ingredient as compared to earlier works is a manifestly 
gauge invariant Hamiltonian approach \cite{TT} which allows to disentangle gauge transformations from 
perturbation theory and thus can be unambiguously applied to any perturbative order. 

This works as follows: In presence of an exact Killing symmetry, the Einstein equations 
including matter often allow for exact solutions. In absence of the exact Killing symmetry 
but close to it, this motivates to separate the degrees into symmetric ``background'' (s) 
and non-symmetric ``perturbations'' (n). In particular the constraints of general relativity (GR)
can be separated into symmetric and non-symmetric. This motivates a second separation 
of the degrees of freedom into ``gauge'' (g) and ``true'' (t). In the Hamiltonian formulation 
we then have 4 sets of canonical pairs with qualifiers st, sg, nt, ng. We then 1. impose 
gauge fixing conditions (G) on the sg and ng configuration degrees of freedom, 2. solve the 
symmetric and non-symmetric constraints (C) respectively for the sg and ng momentum degrees of 
freedom and 3. solve the stability conditions (S) on the symmetric and non-symmetric 
smearing functions of the symmetric and non-symmetric constraints respectively for the 
physical lapse and shift functions. This 
then defines the central ingredient of this framework, the reduced Hamiltonian H which drives
the dynamics of the remaining st and nt degrees of freedom. All of 
this can be done non-perturbatively but generically only in implicit form which makes this unpractical. 
To obtain a practically useful explicit form we must resort to perturbation theory. Thus the   
solutions to the above 3 conditions G, C, S are expanded with respect to the remaining 
st degrees of freedom which are considered first order. By assumption, the zeroth order,
equivalent to solving the Einstein equations in the presence of the exact Killing symmetry, 
can be solved exactly which therefore enables to solve all equations G, C, S to any desired
order and therefore to expand H to any desired order. Note that this so-called reduced 
phase space formulation retains both st and nt degrees of freedom as dynamical variables and 
allows upon quantisation for backreaction, not only between quantum matter and 
quantum geometry but also between the dynamical ``quantum background''
and dynamical ``quantum perturbations'' and therefore is in spirit very similar to the 
hybrid framework of \cite{Gomar} developed for cosmological perturbations.   

The most important Killing symmetries for GR are those associated with homogeneous, spherically
symmetric and axi-symmetric spacetimes giving rise to the well known cosmological and 
black hole solutions. In \cite{TT} the above manifestly gauge invariant formalism was already 
tested and compared to \cite{Gomar} where exact match was found at second order. In this 
paper and the next \cite{III} of this series we consider spherically symmetric black hole 
perturbations for vacuum GR and Einstein-Maxwell theory respectively. We develop the general 
theory sketched above explicitly for this situation and compute the reduced Hamiltonian 
to second order in the nt perturbations. As motivated in \cite{I}, in order to explore the 
interior of the black hole and to be able to discuss the possibility of singularity resolution and 
black hole -- white hole transitions (BHWHT) it is important to pick the gauge fixing condition to be 
compatible with coordinate charts covering also the black hole and/or white hole interior. 
This requirement selects the outgoing and ingoing so-called Gullstrand -- Painlev\'e gauge (GPG) 
which describes
in spherically symmetric vacuum spacetimes a geodesic congruence of timelike observers whose
simultaneity surfaces foliate the spacetime including the white and black hole respectively. 
In these coordinates the zeroth order 
spatial geometry is exactly flat and the spacetime geometry is asymptotically 
flat which makes this gauge extremely convenient also for purposes of QFT in curved 
spacetime (CST) applications which become relevant when quantising the reduced Hamiltonian
whose second order selects Fock representations that are used for the perturbative treatment 
of the higher orders.

Therefore our second order reduced Hamiltonian has to be 
translated from GP coordinates into the usual Schwarzschild coordinates in the asymptotic 
regions in order to compare it with the Hamiltonian derived in \cite{6} which 
was shown to reproduce the seminal Regge-Wheeler \cite{1} and Zerilli \cite{2} equations for linear 
axial (or ``odd'') and polar (or ``even'') gravitational and electromagnetic perturbations. 
We show that after a suitable combination of canonical transformations we expectedly 
find exact match with those works. In another paper of this series \cite{IV} we also 
consider purely geometrical perturbations in generalised gauges in order to 
compare our formalism 
with the general gauge formulations of \cite{3,4,5,7,8}.\\
\\
This paper is organised  as follows. \\
\\

In Section \ref{sec:Outline} we briefly review the perturbative 
techniques outlined in \cite{I}. 
Next, in Section \ref{sec:perturbHamiltonian} we expand the constraints of general relativity to 
second order in the perturbations. Then in Section \ref{sec:Analysis} we review and expand the 
reduced Hamiltonian to second order  
and simplify it using canonical transformations. In Section \ref{sec:Comparison} we compare the results to the 
ones obtained in the Lagrangian formulation. In the appendices we review the definition of the spherical 
harmonics and provide some details on the expansion of the constraints to second order in the perturbations. Finally, we review some aspects of the Lagrangian perturbation theory for spherical symmetry. 

\section{Review of the reduced phase space approach adapted to spherical symmetry}
\label{sec:Outline}

In this section we detail the general discussion of \cite{I}. In the first subsection 
introduce our notation applied to the Hamiltonian formulation of general relativity in terms of 
ADM variables. 
In the second subsection, we explain the reduced phase space construction using a canonical 
chart adapted to spherical symmetry.

\subsection{Hamiltonian Formulation of General Relativity}

For the Hamiltonian formulation of general relativity we follow the ADM approach \cite{Arnowitt1962}. 
We assume spacetime to be globally hyperbolic and foliated into three dimensional hypersurfaces diffeomorphic to 
$\sigma$. The topology of spacetime is therefore given by $\mathbb{R} \times \sigma$. 
We choose coordinates $(t,\bm x)$ adapted to the foliation and use indices $\mu,\nu,\dots=1,2,3$ for quantities on $\sigma$. 

Explicitly, we will split the metric according to
\begin{equation}
    \dd{s}^2 = - (N^2 - m_{\mu \nu} N^\mu N^\nu) \dd{t}^2 + 2 m_{\mu\nu} N^\mu \dd{t}\dd{\bm x}^\nu + m_{\mu \nu} \dd{\bm x}^\mu \dd{\bm x}^\nu\,,
\end{equation}
where $N$ is the lapse function, $N^\mu$ is the shift vector and $m_{\mu \nu}$ is the induced metric on the hypersurfaces of the foliation. 

Starting from the Einstein-Hilbert action we derive the corresponding Hamiltonian theory. As the Legendre 
transformation is singular, we need to use the Dirac algorithm. The Hamiltonian of the theory is fully 
constrained. It is a linear combination of constraints and reads
\begin{equation}
    H = \int \dd{\Sigma} \qty(N V_0 + N^\mu V_\mu)\,,
\end{equation}
where $N$ and $N^\mu$ enter as Lagrange multipliers for the constraints $V_0$ and $V_\mu$. 
These constraints depend on the variable $m_{\mu \nu}$ and its conjugate momentum $W^{\mu \nu}$. 
$V_0$ is called the Hamiltonian constraint and given by
\begin{equation}
    V_0 = \frac{1}{\sqrt{\det(m)}}\qty(m_{\mu \rho} m_{\nu \sigma} - \frac{1}{2} m_{\mu \nu} m_{\rho \sigma})W^{\mu \nu}W^{\rho \sigma}- \sqrt{\det(m)} R \,.
\end{equation}
where $R$ is the Ricci scalar of the induced metric $m_{\mu \nu}$. $V_\mu$ is the diffeomorphism constraint and is defined as
\begin{equation}
    V_\mu = - 2 m_{\mu \rho} D_\nu W^{\nu \rho}.
\end{equation}
The conjugate variables $m_{\mu \nu}$ and $W^{\mu \nu}$ satisfy the Poisson brackets
\begin{equation}
    \qty{m_{\mu \nu}(t,\bm x),W^{\rho \sigma}(t,\bm y)} = \delta^\rho_{(\mu} \delta^\sigma_{\nu)} \delta(\bm x,\bm y)
\end{equation}

\subsection{The reduced phase approach to Hamiltonian theories with constraints adapted to spherical symmetry}

\

We follow the general exposition in \cite{TT,I} and provide more details for the concrete application 
to spherical symmetry. \\
\\
The canonical variables $(m_{\mu\nu}, W^{\mu\nu})$ and 
the constraints $(V_\mu,V_0)$ 
are spatial tensor fields of density weight zero and one respectively. Likewise 
the smearing functions $S^\mu:=N^\mu,\; S^0:=N$ of the constraints are spatial tensor fields of 
density weight zero. These transform in corresponding representations of the 
rotation subgroup of the spatial diffeomorphism group SO(3) which in turn can be decomposed into irreducible 
representations labelled by $l=0,1,2,..$ of dimension $2l+1$. If we denote by $\mu=3$ 
the radial direction and $\mu=A,B,..=1,2$ the angular direction then $(m_{33},W^{33}),\;
(V_0,S^0)$ can be decomposed 
into scalar (``vertical'') 
spherical harmonics $L_{lm}=:L_{v,l,m},\; |m|\le l\ge 0$. Next, $(m_{3A},W^{3A}),\;(V_A, S^A)$ 
can be decomposed 
into polar (``even'') and axial (``odd'') vector spherical harmonics $L{\alpha,l,m}_A,\; |m|\le l\ge 1,\;
\alpha=e,o$. Then 
$(m_{AB},W^{AB})$ can be decomposed into polar (``horizontal'' and ``even'') 
and axial (``odd'')
tensor spherical harmonics,
$L_{h,l,m}^{AB},\; |m|\le l\ge 0$ and $L{\alpha,l,m}_{AB},\; |m|\le l\ge 2,\;
\alpha=e,o$. Finally $V_3,S^3$ can be decomposed into scalar (``horizontal'') 
spherical harmonics $L_{lm}=:L_{h,l,m},\; |m|\le l\ge 0$.
See \cite{I} for an explanation for this particular terminology. More 
details about these harmonics will be given below and it is understood that spherical 
indices are moved with the round metric $\Omega_{AB}$ on the sphere. 

When present, we single out the ``symmetric'' (i.e. $l=0$) mode of a configuration variable, 
momentum, constraint 
or smearing function and denote it by $q,\;p,\;C,\;f$ respectively. These carry 
labels $\alpha=v,h,e,o$ only. The non-symmetric (i.e. 
$l>0$) scalar and vector modes will be denoted by $x,y,Z,g$ while the non-symmetric 
tensor modes will be denoted by $X,Y$. These carry in addition $l,m$ labels.
Note that automatically $q^\alpha=p_\alpha=C_\alpha=f^\alpha=0$ for $\alpha=e,o$ and 
$X^{\alpha,l,m}=Y_{\alpha,l,m}=0$ for $\alpha=v,h$. In this way there is 
precisely one symmetric canonical pair for each symmetric constraint and 
precisely one non-symmetric canonical pair for each non-symmetric constraint. This 
suggests to solve the symmetric constraints $C$ for the symmetric momenta $p$ and 
the non-symmetric constraints $Z$ for the non-symmetric momenta $y$. Then it is natural 
to impose gauge fixing conditions on the corresponding configuration coordinates 
$q,x$. The degrees of freedom $(X,Y)$ are left untouched. If there was additional 
matter such as the Klein-Gordon field considered in \cite{I} then there would be additional
symmetric degrees of freedom $(Q,P)$ (and additional non-symmetric ones extending the 
list of the $(X,Y)$). The $(Q,P), (X,Y)$ are therefore identified as the observable 
or true degrees of freedom. Altogether we obtain a canonical chart made out of four sets of canonical pairs displayed 
in table \ref{tab:variables}.

\begin{table}[h!]
    \centering
    \begin{tabular}{c c c}
    \toprule
    & gauge d.o.f. & true d.o.f.\\
    \midrule 
    symmetric & 
    $(q,p)$ & $(Q, P)$\\
    non-symmetric & 
    $(x, y)$ & $(X, Y)$ \\
    \bottomrule
    \end{tabular}
    \caption{Decomposition of the variables into symmetric vs. non-symmetric and gauge vs. true degrees of freedom}
    \label{tab:variables}
\end{table}

Notice that upon imposing the gauge fixing condition on $q,x$ and solving the constraints for
$(p,y)$ the variables $q,x$ have disappeared while $p,y$ are now functions of $(Q,P,X,Y)$. 
These exact functions can be obtained non-perturbatively but only in implicit form as shown in \cite{I}.
They can be used to find the reduced Hamiltonian and the solution of the stability conditions 
also in exact but implicit form \cite{I}. In this way the step of extracting the 
full gauge invariant information of the theory can be considered as being completed and there 
are no constraints left over of which we need to verify the first class consistency property
(the unreduced constraints are first class). 

To obtain an explicit expressions we now invoke perturbation theory with respect to 
the true non-symmetric degrees of freedom $X,Y$. To do this one expands 
the solutions 
$p=p^\ast,y=y^\ast$ of the constraints $C=Z=0$ with the gauge $q=q_\ast, x=x_\ast$ imposed
in powers of $X,Y$, say $p^\ast=\sum_{n=0}^\infty\; p^\ast(n)$ and similar for $y^\ast$ where 
$p^\ast(n), y^\ast(n)$ are of $n-th$ order in $X,Y$. Then we expand the condition 
$C_\ast(Q,P,X,Y):=C(q=q_\ast,p=p^\ast,x=x_\ast,y=y^\ast,Q,P,X,Y)=0$ into powers of $X,Y$ and similar for 
$Z$. We isolate the $n-th$ order contributions, say $C_\ast=\sum_{n=0}^\infty\; C_\ast(n)$ 
and similar for $Z$. At this point the adaption of the canonical chart to the symmetry group 
becomes crucial: By construction one finds $C_\ast(1)=Z_\ast(0)\equiv 0$ and this 
leads automatically to the condition $p_\ast(1)=y_\ast(0)=0$. Then, since by assumption
$C_\ast(0)=0$ is exactly solvable for $p_\ast(0)$ one finds that the remaining 
equations $Z_\ast(1)=0, \;C_\ast(2)=0, Z_\ast(2)=0,\;.., C_\ast(n)=0,\; Z_\ast(n)=0,..$ 
can be consecutively solved for $y^\ast(1),\; p^\ast(2),\; y^\ast(2),..,p^\ast(n), y^\ast(n)$.
In particular, to find $p^\ast(2)$ which will be of central interest for the present paper,
we just need to construct $y^\ast(1)$. See \cite{TT} for the detailed proof.

\section{Perturbative expansion of the constraints}
\label{sec:perturbHamiltonian}

In this section we present the perturbation theory around a spherically symmetric spacetime in the Gullstrand 
Painlevé (GP) gauge. We consider the theory in the Hamiltonian setup. Similar works have been done 
before in \cite{6,7,8}. In this section we follow these works closely with 
some generalizations. In the work \cite{6} Moncrief discusses perturbations in the exterior 
region of the black hole in the Schwarzschild gauge. Our formalism uses the Gullstrand Painlevé gauge instead 
because they are regular 
at the horizon and can be used to explore the black hole interior which is a crucial property as we have 
quantum applications in mind towards the question of black hole evaporation, see \cite{I}. 

The works by Brizuela and Martín-García \cite{7,8} consider the perturbation 
theory without specifying the gauge and thus their analysis could be specialised to our GPG.
However, their approach does not follow the manifestly gauge invariant reduced phase space 
route advertised in this series of works which is applicable at any order. Since we want to test 
our approach and compare it at second order with the results of other approaches such as 
\cite{7,8}, we strictly follow our formalism sketched in the previous section
and outlined in detail in \cite{TT,I}. In the course of our analysis we demonstrate 
that it is possible to simplify the reduced Hamiltonian in the Gullstrand Painlev\'e gauge using 
suitable canonical transformations.
The correspondence between the physical Hamiltonian and the Lagrangian formulation 
(see Appendix \ref{sec:PerturbLagrangian}) is also very clear in the corresponding canonical 
chart.\\

The outline of this section is as follows. 
First, we recall the parametrisation of the spherically symmetric ``background ''variables $(q,p)$ 
and the corresponding symmetric zeroth order constraints $C(0)$ with respect to the gravitational
perturbations) following the notation of \cite{I}. 
We explicitly solve $C(0)=0$ in the Gullstrand-Painlevé gauge which is the zeroth 
order step in the construction algorithm sketched in the previous section. 
In the second subsection we parametrise the gravitational perturbations of this background 
and expand all the constraints to second order in the perturbation variables. More precisely 
it is sufficient to compute the symmetric constraints $C=C(0)+C(2)+..$ to second order and the 
non-symmetric constraints $Z=Z(1)+..$ to first order.

\subsection{The spherically symmetric background}

Spherically symmetric spacetimes in the Hamiltonian formulation are studied extensively in the literature 
(see e.g. \cite{Benguria1977,Kuchar1994}). 
We review this formalism and recover the basic results using the 
reduced phase space formalism. 
This prepares and sets up the notation for the perturbation theory in the next sections. 

In the Hamiltonian formulation of general relativity we have the foliation of spacetime with topology 
$\mathbb{R} \times \sigma$. For spherical symmetry we further consider a foliation of the spatial slices as 
$\sigma = \mathbb{R}_+ \times S^2$ where $S^2$ is the 2-sphere, 
if we have one asymptotic end. If we have several asymptotic 
ends such as the black hole white hole transition spacetime discussed in \cite{I} we have 
correspondingly several copies of $\mathbb{R}_+$ together with gluing 
prescriptions for the 
overlapping coordinates charts. The following analysis applies to each asymptotic
end separately as the physical Hamiltonian is a boundary term (there is no 
boundary in the overlapping  
coordinate charts). It will therefore be sufficient to consider one asymptotic end only for what 
follows so that the radial variable takes positive values only. 

On the two sphere we consider standard spherical coordinates $(\theta,\phi)$ and for the indices we use 
capital letters $(A,B,C, \dots=1,2)$. The metric on $S^2$ is denoted by $\Omega_{AB}$ and the unique 
torsion-free connection compatible with $\Omega_{AB}$ is $D_A$. For the
radial coordinate we use the notation $r$ and the index $3$.

It is well known that in this decomposition the most general form of a spherically 
symmetric metric is given by
\begin{equation}
    m_{33} = e^{2\mu}, \quad \quad m_{3A} = 0, \quad \quad m_{AB} = e^{2\lambda} \Omega_{AB}\,.
\end{equation}
The only degrees of freedom are $\mu$ and $\lambda$. They are arbitrary functions of time 
$\tau$ and $r$. We use here an a priori arbitrary time coordinate $\tau$ and reserve 
the label $t$ for the Schwarzschild gauge.

The form of the conjugate momenta $W^{\mu \nu}$ is easily read off from the symplectic term 
$\int \dd[3]{x} \dot{m}_{\mu \nu}W^{\mu \nu}$. In order that $(\mu,\pi_\mu)$ and $(\lambda, \pi_\lambda)$ 
form symplectic pairs we have 

\begin{equation}
    W^{33} = \sqrt{\Omega} \frac{\pi_\mu}{2}e^{-2\mu}, \quad \quad W^{3A} = 0, \quad \quad W^{AB} = \sqrt{\Omega} \frac{\pi_\lambda}{4}e^{-2\lambda} \Omega^{AB}\,,
\end{equation}
where $\sqrt{\Omega} = \sqrt{\det \Omega}$. 

In the literature several gauge choices for the spherically symmetric variables are discussed. 
Most of them include the area gauge which sets $\lambda = \log(r)$. 
For the Schwarzschild gauge we further restrict to $\pi_\mu = 0$. 
As it turns out this gauge choice leads to an unphysical divergence of the function $\mu$ 
at the event horizon of the black hole consistent 
with the fact that Schwarzschild coordinates only cover the exterior of the spacetime.
A choice of gauge which avoids this problem is the GP gauge which imposes $\mu = 0$. 
For that reason we will use the GP gauge in the following. For the many other reasons 
to prefer that gauge and its generalised version, see the general discussion in \cite{I}.

We insert this form of metric and its conjugate momentum into the expressions for the constraints. 
The result is independent of the angular components and we can integrate them out. 
We obtain the following list of symmetry reduced constraints, following the notation 
of the previous section
\begin{align}
    {}^{(0)}C_v &=\frac{4 \pi}{8 r^2} \qty(\pi_\mu^2 - 2 \pi_\mu \pi_\lambda)\,,\\
    {}^{(0)}C_h &= 4 \pi \qty(\frac{\pi_\lambda}{r} - \pi_\mu')\,.
    \label{eq:BackCv}
\end{align}
In these equations and in the following we denote the radial derivative $\partial_r$ by a prime $'$. 

The next step is the solution of these constraints for the momenta.
From the radial diffeomorphism constraint we obtain
\begin{equation}
    \pi_\lambda = r \pi_\mu'
\end{equation}
Inserting this expression into ${}^{(0)}C_v$ we obtain the differential equation
\begin{equation}
    \partial_r \qty(\frac{(\pi_\mu)^2}{r}) = 0
\end{equation}
The solution of this differential equation is straight forward and given by
\begin{equation}
    \pi_\mu^2 = 16 r r_s\,,
    \label{eq:pimusquare}
\end{equation}
where $r_s$ is an integration constant. Equation \eqref{eq:BackCv} is easily solved for $\pi_\lambda$ and we have
\begin{equation}
    \pi_\lambda = \frac{1}{2}\pi_\mu\,.
\end{equation}

It remains to physically interpret the integration constant $r_s$. As the name suggests it is precisely the Schwarzschild radius. To see this, we compare the extrinsic curvature calculated in two different ways. For our ansatz for the metric and its conjugate momentum we have that the only non-vanishing components are
\begin{equation}
    K^{33} = \frac{1}{4r^2 }\qty(\pi_\mu - \pi_\lambda)\,, \quad \quad K^{AB} = -\frac{1}{4 r^4}\Omega^{AB}\pi_\mu\,.
\end{equation}
For the metric in GP coordinates the extrinsic curvature takes the form $K^{33} = \frac{\sqrt{r r_s}}{2r^2}$ and $K^{AB} = - \frac{\sqrt{r_s r}}{r^4}\Omega^{AB}$. Inserting the expressions for $\pi_\mu$ and $\pi_\lambda$ we get exact agreement provided that for the square root of equation \eqref{eq:pimusquare} we take the positive sign. In summary we have the solution
\begin{align}
    \pi_\mu &= 4\sqrt{r r_s}\\
    \pi_\lambda &= 2\sqrt{r r_s}
\end{align}

The results of this subsection will be directly relevant for the perturbative expansion of the 
reduced Hamiltonian by the iterative scheme sketched in the previous section as they encode 
precisely the solution of the corresponding zeroth order equations. Namely the above 
solutions for $\pi_\mu, \pi_\lambda$ are precisely the zeroth 
order solutions $\pi_\mu(0), \pi_\lambda(0)$ of the 
symmetric constraints.

\subsection{Gravitational Perturbations}

We now include the non-symmetric perturbations.
It is convenient to expand the full metric $m$ and its conjugate momentum $W$ in terms of tensor spherical 
harmonics. The definitions that we will use as well as some useful formulae are recorded in appendix 
\ref{sec:sphericalHarmonics}. Then in the GPG we have the following expansions:
\begin{align}
    m_{33} &= 1 +  \sum_{l\geq 1, m} \bm x^v_{lm} L_{lm}\\
    m_{3A} &= 0 + \sum_{l \geq 1, m,I} \bm x^I_{lm} L^{I,lm}_A\\
    m_{AB} &= r^2 \Omega_{AB} + \sum_{l\geq 1,m} \bm x^h_{lm} \Omega_{AB} L_{lm} + \sum_{l \geq 2, m, I} \bm X^I_{lm} L^{I,lm}_{AB}\\
    W^{33} &= \sqrt{\Omega}\frac{\pi_\mu}{2} + \sqrt{\Omega} \sum_{l\geq 1, m} \bm y_v^{lm} L_{lm}\\
    W^{3A} &= 0 +  \sqrt{\Omega} \frac{1}{2}\sum_{l \geq 1, m,I}\bm y_I^{lm} L_{I,lm}^A\\
    W^{AB} &= \sqrt{\Omega} \frac{\pi_\lambda}{4 r^2} \Omega^{AB} + \sqrt{\Omega} \frac{1}{2}\sum_{l\geq 1,m} 
    \bm y_h^{lm} \Omega^{AB} L_{lm} + \sqrt{\Omega} \sum_{l \geq 2, m, I} \bm Y_I^{lm} L_{I,lm}^{AB}
\end{align}

The various functions $L$ are the spherical scalar, vector and tensor harmonics. 
They are defined in appendix \ref{sec:sphericalHarmonics} normalised such that they 
form orthonormal bases on the corresponding Hilbert spaces \cite{1}. The restrictions 
on the ranges of $l$ displayed follow from the corresponding completeness relations.
The label $I$ runs over the set $\{e,o\}$ labeling the even and odd vector and tensor harmonics.
We distinguish it from the label $\alpha$ of the previous section which also includes the
$v,h$ labels the dependence on which we have instead split off in the above decomposition. 

Later, we will see that the physical Hamiltonian is given by a boundary term. In the computation it is therefore necessary to carefully analyse the behaviour of the perturbations close to the boundary. In our approach we are considering the boundary as $r$ approaches infinity and need to select appropriate fall off conditions there. These conditions need to be sufficiently general to allow for interesting physical solutions. 
In the first paper of the series \cite{I} a specific choice for the 
boundary conditions is proposed which is adapted 
to the GP gauge. In the following we recall them in the present 
notation. For that we call the perturbations to the metric $\delta m_{\mu \nu}$ and the perturbations to the momenta $\delta W^{\mu \nu}$. Then, we have
\begin{align}
    \label{eq:Decay}
    \begin{split}
    \delta m_{33} &\sim \delta m_{33}^+ r^{-1} + \delta m_{33}^- r^{-2}\\
    \delta m_{3A} &\sim \delta m_{3A}^+ + \delta m_{3A}^- r^{-1}\\
    \delta m_{AB} &\sim \delta m_{AB}^+ r + \delta m_{AB}^-\\
    \delta W^{33} &\sim \delta W^{33}_- + \delta W^{33}_+ r^{-1}\\
    \delta W^{3A} &\sim \delta W^{3A}_- r^{-1} + \delta W^{3A}_+ r^{-2}\\
    \delta W^{AB} &\sim \delta W^{AB}_- r^{-2} + \delta W^{AB}_+ r^{-3}\,.
    \end{split}
\end{align}
On the right-hand side, $\delta m_{\mu \nu}^\pm$ and $\delta W^{\mu \nu}_\pm$ only depend on the angular coordinates $\theta,\phi$ and are constant with respect to the radial coordinate $r$. 
We introduced the notation $+$/$-$ to represent even/odd behaviour under the parity transformation $\hat P$ respectively. This is different from the conventions in \cite{I} in order to avoid confusion with the labels even and odd of the vector and tensor spherical harmonics. It is important to note that the notions of even and odd parity are different from the terms even and odd used to classify vector and tensor harmonics. In fact, in the literature the vector and tensor harmonics are sometimes split into axial (odd) and polar (even) modes. The even modes transform in the same way as the scalar spherical harmonics ($\hat P = (-1)^l$) and the odd modes transform with an additional sign ($\hat P =(-1)^{l+1}$).

In the variables $x,y,X,Y$ introduced above we expand the non-symmetric constraints to first and the symmetric constraints to second order. The zeroth order non-symmetric constraints as well as the first order symmetric constraints vanish identically. Some details and intermediate steps of the computation are documented in appendix \ref{sec:Expansion2ndOrder}

We obtain the following list of non-symmetric first order constraints:
\begin{align}
    {}^{(1)}Z^h_{lm} &= -2 \partial_r \bm y_v + \sqrt{l(l+1)} \bm y_e + 2 r \bm y_h - \partial_r \pi_\mu \bm x^v - \frac{1}{2} \pi_\mu \partial_r \bm x^v +\frac{\pi_\lambda}{2r^2} \sqrt{l(l+1)} \bm x^e + \frac{\pi_\lambda}{2 r^2} \partial_r \bm x^h\\
    {}^{(1)}Z^e_{lm} &= \sqrt{2(l+2)(l-1)}\qty(r^2 \bm Y_e  + \frac{\pi_\lambda}{4 r^2}\bm X^e) -\partial_r(r^2 \bm y_e + \pi_\mu \bm x^e)  - \sqrt{l(l+1)} r^2 \bm y_h + \sqrt{l(l+1)} \frac{\pi_\mu}{2} \bm x^v\\
    {}^{(1)}Z^o_{lm} &= \sqrt{2(l+2)(l-1)}\qty( r^2 \bm Y_o + \frac{\pi_\lambda}{4r^2} \bm X^o) - \partial_r \qty(r^2 \bm y_o + \pi_\mu \bm x^o)\\
    {}^{(1)}Z^v_{lm} &=  \frac{1}{2 r^2} (\pi_\mu - \pi_\lambda) \bm y_v- \frac{1}{2} \pi_\mu \bm y_h + 2 \qty(\partial_r^2 - \frac{1}{r}\partial_r -\frac{(l+2)(l-1)}{2 r^2} - \frac{r_s}{r^3}) \bm x^h\\
    &-\qty(2 r \partial_r + l(l+1) + 2 - 2 \frac{r_s}{r}) \bm x^v + 2 \qty(\partial_r + \frac{1}{r}) \sqrt{l(l+1)} \bm x^e -\sqrt{\frac{(l+2)(l+1)l(l-1)}{2}} \frac{1}{r^2} \bm X^e 
\end{align}
For simplicity we did not display the labels of the spherical harmonics.
 
The second order symmetric constraints 
involve a sum over all the labels $l,m$. 
Every quadratic term in the following equations has an implicit sum over $l$ and $m$. 
Explicitly we abbreviated the expression using for example
\begin{equation}
    \bm x^o \cdot \partial_r \bm y^o := \sum_{lm} \bm x^o_{lm} \partial_r \bm y^o_{lm}\,.
\end{equation}
In this notation we find after integrating out the angle dependence\\
\begin{align}
    {}^{(2)}C_h &= - \bm x^o \cdot \partial_r \bm y_o + \bm Y_o \cdot \partial_r \bm X^o + \bm y_v \cdot \partial_r \bm x^v - 2 \partial_r (\bm x^v \cdot \bm y_v) - \bm x^e \cdot \partial_r \bm y_e + \bm Y_e \cdot \partial_r \bm X_e + \bm y_h \cdot \partial_r \bm x^h \\
    {}^{(2)}C_v &= \frac{1}{2} \bm y_o \cdot \bm y_o + \frac{1}{2 r^2}\pi_\mu \bm x^o \cdot \bm y_o + \frac{1}{r^2}\bm x^o \cdot \qty(4 r \partial_r + \frac{l(l+1)}{2} - 3 + 2 \frac{r_s}{r})\bm x^o + r^2 \bm Y_o \cdot \bm Y_o  - \frac{1}{4 r^2}\pi_\mu \bm Y_o \cdot \bm X^o\nonumber\\
    &-\frac{1}{r^4} \bm X^o \cdot \qty(r^2 \partial_r^2 - 4 r \partial_r + \frac{7}{2} - \frac{r_s}{4 r})\bm X^o - \frac{3}{4r^2}\partial_r \bm X^o \cdot \partial_r \bm X^o -\frac{1}{r^3} \sqrt{\frac{(l+2)(l-1)}{2}}\bm x^o \cdot \qty( r \partial_r -2) \bm X^o\nonumber\\
    &+\frac{1}{2r^2} \bm y_v \cdot \bm y_v + \frac{5}{8 r^2} \pi_\mu \bm x^v \cdot \bm y_v + \bm x^v \cdot \qty(3 r \partial_r + 1+\frac{r_s}{r})\bm x^v + \frac{1}{2} \bm y_e \cdot \bm y_e + \frac{1}{2 r^2} \pi_\mu \bm x^e \cdot \bm y_e\nonumber\\
    &+ \frac{1}{r^2} \bm x^e \cdot \qty(4r \partial_r - 3 + 2\frac{r_s}{r})\bm x^e - \frac{1}{2 r^4} \qty(- 4 r \bm x^h \cdot \partial_r \bm x^h + r^2 \partial_r \bm x^h \cdot \partial_r \bm x^h + 4 \qty(1 - \frac{r_s}{r}) \bm x^h \cdot \bm x^h)\\
    &+ r^2 \bm Y_e \cdot \bm Y_e - \frac{1}{4r^2}\pi_\mu \bm X^e \cdot \bm Y_e + \frac{1}{4r^4} \qty( - 3 r^2 \partial_r \bm X^e \cdot \partial_r \bm X^e + \bm X^e \cdot \qty(- 4r^2 \partial_r^2 + 16 r \partial_r - 14 + \frac{r_s}{r})\bm X^e)\nonumber\\
    & - \bm y_h \cdot \bm y_v - \frac{\pi_\mu}{2r^4}\bm y_v \cdot \bm x^h - \frac{1}{4}\pi_\mu \bm x^v \cdot \bm y_h - \frac{3 r_s}{r^3} \bm x^v \cdot \bm x^h - \bm x^v \cdot \qty(\frac{1}{2r^2} l(l+1) \bm x^h + \partial_r^2 \bm x^h + \frac{1}{r^2} \bm x^h - \frac{1}{r} \partial_r \bm x^h)\nonumber\\
    &- \partial_r \bm x^h \cdot \partial_r \bm x^v - \frac{1}{2r^2} \sqrt{\frac{(l+2)(l+1)l(l-1)}{2}}\bm x^v \cdot \bm X^e -  \sqrt{l(l+1)} \bm x^e \cdot \qty(\partial_r \bm x^v - \frac{1}{r} \bm x^v)\nonumber \\
    &- \sqrt{l(l+1)}\bm x^v \cdot \partial_r \bm x^e + \frac{1}{r^2}\sqrt{l(l+1)} \bm x^e \cdot \qty(2\frac{1}{r} \bm x^h - \partial_r \bm x^h) + \sqrt{\frac{(l+2)(l-1)}{2}} \frac{1}{r^3} \bm x^e \cdot \qty( 2 \bm X^e - r \partial_r \bm X^e)\nonumber
\end{align}
This expression has been found analytically and has been checked by computer algebra. On the computer we used Mathematica with the xAct package \cite{xAct}. It provides tools to manipulate tensor and spherical tensor harmonics.
Of course it simplifies drastically in the GPG for which $x^{\alpha,l,m}:=0$ but since the equations 
remain managable even before imposing the GPG we keep them in place.

\section{Application of the symplectic reduction formalism to second order}
\label{sec:Analysis}

In this section we compute $p_v, p_h$ with 
$W^{33}/\sqrt{\det(\Omega)}=p_v+y_v,\;
\Omega_{AB}\;W^{AB}/\sqrt{\det(\Omega)}=p_h+y_h$ or equivalently 
$\pi_\mu,\;\pi_\lambda$
to second order in $X,Y,x$ by solving the second order 
symmetric constraints $C_v,C_h$. This requires the solution to first order of $y_v, y_h, y_e, y_o$
using the first order non-symmetric constraints $Z_v, Z_h, Z_e, Z_o$. 

The reduced Hamiltonian derived in \cite{I} is a pure boundary term and depends only on $p_v$. 
To see this requires a careful definition of the phase space and the decay behaviour of the 
canonical variables at infinity together with a boundary term analysis. The boundary
terms ensure that the constraints are functionally differentiable and they look different 
than in the standard Schwarzschild coordinates because the information about the black hole 
mass now resides in the momenta conjugate to the three metric rather than the three metric 
(which in GPG is by construction exactly flat). This requires a careful transcription 
of the usual field asymptotics by performing the corresponding boost between the two systems 
of coordinates.

In the first subsection we compute $p_\alpha(2), \;\alpha=v,o$ which 
relies on expressions for $p_\alpha(0), y_\alpha(1)$. We already computed 
$p_\alpha(0)$ while 
$y_\alpha(1),\;\alpha=v,h,e,o$ will be computed in the second section 
which just requires the already 
known solutions $p_\alpha(0)$. The reason for why we compute $p_\alpha(2)$ first 
is that we can solve the required radial differential equations without knowing 
the explicit form of $y_\alpha(1)$.  As in earlier treatments 
the equations for $y_o, y_e$ decouple so that these cases can be treated 
separately. 
Finally we recall from \cite{I} the reduced Hamiltonian and expand it to second order 
using the now known terms $p_v(0), p_v(2)$ .

\subsection{Solution of the second order constraints}

Before, we already solved the zeroth order constraints explicitly. 
Furthermore, we assume that we successfully solved the first order. 
Given this, we investigate the second order constraints in this section. 
First, we consider a splitting of the background momenta as 
$\pi_\mu^{(0)} + \pi_\mu^{(2)}$ and $\pi_\lambda^{(0)} + \pi_\lambda^{(2)}$. 
The expressions for $\pi_\mu^{(0)}$ and $\pi_\lambda^{(0)}$ were already obtained 
in the discussion of the spherically symmetric sector. 
The second order corrections $\pi_\mu^{(2)}$ and $\pi_\lambda^{(2)}$ are functions of second 
order in the perturbations which we determine in the following. 

We insert the expressions for $\pi_\mu$ and $\pi_\lambda$ into the full constraints and truncate at second order. We obtain
\begin{align}
    C_v &\sim \frac{4 \pi}{4r^2} \qty(\pi_\mu^{(0)}\pi_\mu^{(2)} - \pi_\mu^{(0)} \pi_\lambda^{(2)} - \pi_\mu^{(2)} \pi_\lambda^{(0)}) + {}^{(2)}C_v=0\,,\\
    C_h &\sim 4 \pi\qty(\frac{1}{r} \pi_\lambda^{(2)} - (\pi_\mu^{(2)})') + {}^{(2)}C_h=0\,.
\end{align}
Note that the zeroth order contributions vanish because $\pi_\mu^{(0)}$ and $\pi_\lambda^{(0)}$ 
are solutions of these equations. The terms $ {}^{(2)}C_v$ and ${}^{(2)}C_h$ are given by
\begin{align}
    {}^{(2)}C_h &= \bm Y_o \cdot \partial_r \bm X^o + \bm Y_e \cdot \partial_r \bm X^e\,,\\
    {}^{(2)}C_v &= \frac{1}{2} \bm y^{(1)}_o \cdot \bm y^{(1)}_o + \frac{1}{2} \bm y^{(1)}_e \cdot \bm y^{(1)}_e +\frac{1}{2r^2} \bm y^{(1)}_v \cdot \bm y^{(1)}_v - \bm y^{(1)}_h \cdot \bm y^{(1)}_v\nonumber\\
    &+ r^2 \bm Y_o \cdot \bm Y_o  - \frac{1}{4 r^2}\pi_\mu^{(0)} \bm Y_o \cdot \bm X^o -\frac{1}{r^4} \bm X^o \cdot \qty(r^2 \partial_r^2 - 4 r \partial_r + \frac{7}{2} - \frac{r_s}{4 r})\bm X^o - \frac{3}{4r^2}\partial_r \bm X^o \cdot \partial_r \bm X^o\\
    &+ r^2 \bm Y_e \cdot \bm Y_e - \frac{1}{4r^2}\pi_\mu^{(0)} \bm Y_e \cdot \bm X^e  - \frac{1}{r^4} \bm X^e \cdot \qty(r^2 \partial_r^2 - 4 r \partial_r + \frac{7}{2} - \frac{r_s}{4 r})\bm X^e - \frac{3}{4r^2}\partial_r \bm X^e \cdot \partial_r \bm X^e\nonumber\,.
\end{align}
These are the second order symmetric constraints where we imposed the GPG gauge $\bm x=0$ and inserted the solution of the first order non-symmetric constraints $\bm y^{(1)}$. The functions ${}^{(2)} C_h$ and ${}^{(2)} C_v$ are quadratic in $\bm y^{(1)}, \bm X, \bm Y$ and they depend on the zeroth order solution $\pi_\mu^{(0)}$ and $\pi_\lambda^{(0)}$.

For the next step we solve the second order constraint equations  
for $\pi_\mu^{(2)}$ and $\pi_\lambda^{(2)}$. From $C_h = 0$ we obtain
\begin{equation}
    \pi_\lambda^{(2)} = r \qty((\pi_\mu^{(2)})' - \frac{1}{4\pi} {}^{(2)}C_h)\,.
\end{equation}
Inserting this into $C_v=0$ we have
\begin{equation}
    \pi_\mu^{(0)} \qty(1 -  \frac{\pi_\lambda^{(0)}}{\pi_\mu^{(0)}})\pi_\mu^{(2)} - r \pi_\mu^{(0)}(\pi_\mu^{(2)})' + \frac{r \pi_\mu^{(0)}}{4\pi} {}^{(2)}C_h + \frac{4 r^2}{4\pi} {}^{(2)}C_v=0\,.
\end{equation}
From the background we can use the radial diffeomorphism constraint to replace $\pi_\lambda^{(0)}$. We obtain

\begin{equation}
    \qty(\frac{1}{r} \pi_\mu^{(0)} \pi_\mu^{(2)})' - \frac{1}{4 \pi r} \pi_\mu^{(0)}{}^{(2)}C_h - \frac{4}{4\pi} {}^{(2)}C_v=0\,.
\end{equation}
This equation is solved by integration and we have
\begin{align}
    \pi_\mu^{(2)} = \frac{4 r}{4 \pi \pi_\mu^{(0)}} \int \dd{r} \frac{\pi_\mu^{(0)} }{4 r} {}^{(2)} C_h + {}^{(2)}C_v\,.
    \label{eq:SolIntegralpimu2}
\end{align}
From $C_v=0$ it is straightforward to get the solution for $\pi_\lambda^{(2)}$ as
\begin{equation}
    \pi_\lambda^{(2)} = \frac{1}{2} \pi_\mu^{(2)} + \frac{4 r^2}{4 \pi \pi_\mu^{(0)}}{}^{(2)}C_v
\end{equation}

Later we use the expansion for $\pi_\mu$ to second order. We have
\begin{align} \label{secondorder}
    \pi_\mu \sim \pi_\mu^{(0)} + \pi_\mu^{(2)} = 4 \sqrt{ r r_s}\qty[ 1 + \frac{1}{16 \pi r_s}
    \int \dd{r}\qty(\sqrt{\frac{r_s}{r}} {}^{(2)}C_h + {}^{(2)}C_v)]
\end{align}
The expression (\ref{secondorder}) displays exactly the phenomenon 
that is responsible for the reduced Hamiltonian to have a chance 
to become a well defined operator in the quantum theory indictaed in 
\cite{I}: Since it is a 
boundary term, it is an object 
not integrated in three but two dimensions only 
which carries the danger to be insufficiently smeared in order that 
the corresponding operator valued distribution becomes an operator. However,
(\ref{secondorder}) automatically carries an additional radial integral when 
evaluating the boundary term on a solution of the gauge, constraint and 
stability conditions.

\subsection{Solution of the first order constraints}

We will solve the first order constraints in three steps. First, we take care of the special case of the dipole perturbations ($l=1$). The reason for this is that the tensor spherical harmonics are not present and therefore, there are no true non-symmetric degrees of freedom. Nonetheless these perturbations have physical significance as they describe how the black hole changes if viewed in an accelerated frame of reference. 

Then, we study the odd parity sector. There is only one constraint $Z^o$ which involves the odd parity variables $\bm x^o$, $\bm y_o$ and $\bm X^o, \bm Y_o$.  We solve this constraint and calculate the second order corrections $\pi_\mu^{(2)}$ explicitly using formula \eqref{eq:SolIntegralpimu2}. The result is further simplified using a canonical transformation.

We complete the analysis of the first order with the study of the even parity constraints. In this sector we study three constraints $Z^v$, $Z^h$ and $Z^e$. According to the program, we need to solve them for $\bm y_v$, $\bm y_h$ and $\bm y_e$. We will end up with a solution to $\pi_\mu^{(2)}$ which only involves the true degrees of freedom $\bm X^e$ and $\bm Y_e$. However, in this section we will not strictly follow the program and instead take a detour. We first solve the constraints for the variables $\bm x^h$, $\bm y_e$ and $\bm Y_e$ and insert the results into the integral for $\pi_\mu^{(2)}$. The result is simplified using some canonical transformations and depends on a pair of true degrees of freedom $Q^e$ and $P_e$. In order to make contact with the GP gauge discussed in this series of papers we then change the gauge and relate the variables $Q^e$ and $P_e$ to the true degrees of freedom  $\bm X^e$ and $\bm Y_e$.

\subsubsection{The dipole perturbations (\texorpdfstring{$l=1$}{l=1})}

We start the analysis of the first order constraints with the the special case $l=1$. In this situation the tensor spherical harmonics $L^{I,lm}_{AB}$ are undefined. Therefore, the non-symmetric true degrees of freedom $\bm X$ and $\bm Y$ are not present. In the analysis we impose the gauge $\bm x^e=\bm x^o=\bm x^h=\bm x^v=0$ and solve the constraints for their conjugate momenta.

For the odd parity degrees of freedom, we obtain from the constraint ${}^{(1)}Z^o_{1m}=0$ the solution
\begin{align}
    \bm y_o^{1 m} = \frac{a_m}{r^2}\,,
\end{align}
with some integration constant $a_m$. This constant is related to the angular momentum of a linearized Kerr solution. To see this, we relate the function $\bm y_o$ to the shift vector in the Lagrangian treatment using the 
stability condition of the gauge fixing $\bm x^o=0$. We have
\begin{align}
    \bm{\dot{x}}^o_{1m} &= \{\bm x^o_{1m}, N {}^{(2)}C_0 + N^3 {}^{(2)} C_r + r^{-2}\delta N^o Z^o\}\Big|_{\bm x^o=0, N = 1, N^3 = \sqrt{\frac{2M}{r}}}\\
    &= \bm y_o^{1m} L^o_A + r^2 \partial_r (r^{-2} \delta N^o_{1m})=0\,.
\end{align}
In this equation we use the solution for the lapse and shift in GP gauge which we will determine in a later section. The solution of the differential equation is 
\begin{equation}
    \delta N^o_{1m} = \frac{a_m}{3r}\,.
\end{equation}
In principle there would be an integration constant. However, the term associated to it would grow quadratically as $r$ approaches infinity. Thus, if the shift vector should decay as $r$ goes to infinity the constant has to be equal to zero. An explanation for why the constant $a_m$ is related to the angular momentum can be found in \cite{5}. 

In the even parity sector there are three first order constraints which we will solve for the three momenta $\bm y_v$, $\bm y_h$ and $\bm y_e$. We start by solving the equation ${}^{(1)} Z^v_{1m}=0$ for $\bm y^{1m}_h$ and obtain
\begin{equation}
    \bm y^{1m}_h = \frac{1}{2r^2} \bm y^{1m}_v\,.
\end{equation}
This relation is used in the equation ${}^{(1)}Z^h_{1m}=0$ and we determine $\bm y^{1m}_e$ in terms of $\bm y^{1m}_v$:
\begin{align}
    \bm y^{1m}_e = \frac{1}{\sqrt{2}}\qty(2 \partial_r \bm y^{1m}_v - \frac{1}{r} \bm y^{1m}_v)\,.
\end{align}
The last remaining constraint equation is ${}^{(1)}Z^e_{1m} = 0$. It gives a differential equation for $\bm y^{1m}_v$:
\begin{align}
    - 2 r^2 \partial_r^2 \bm y^{1m}_v - 3 r \partial_r \bm y^{1m}_v=0\,.
\end{align}
The solution of this differential equation is
\begin{align}
    \bm y^{1m}_v = C^m_1 + \frac{C^m_2}{\sqrt{r}}\,,
\end{align}
with two integration constants $C^m_1$ and $C^m_2$. This equation now fully determines the rest of the momenta:
\begin{align}
    \bm y^{1m}_h &= \frac{1}{2r^2}\qty(C^m_1 + \frac{C^m_2}{\sqrt{r}})\,,\\
    \bm y^{1m}_e &= - \frac{1}{\sqrt{2}}\qty(\frac{C^m_1}{r} + \frac{2 C^m_2}{\sqrt{r}^3})\,.
\end{align}
For the physical interpretation we proceed as in the odd parity case. The stability condition for the gauge fixing $\bm x^o = \bm x^h = \bm x^v=0$ are
\begin{align}
    \dot{\bm x}^e_{1m} &= \{\bm x^e_{1m}, N {}^{(2)}C_0 + N^3 {}^{(2)} C_r + r^{-2}\delta N^e Z^e + \delta N^3 Z^h + \delta N Z^v\}\Big|_{\bm x^o = \bm x^h = \bm x^v=0, N = 1, N^3 = \sqrt{\frac{2M}{r}}}\\
    &=\sqrt{2} \delta N^3_{1m} + r^2 \partial_r (r^{-2}\delta N^e_{1m}) + \bm y_e^{1m}=0\\
    \dot{\bm x}^h_{1m} &= \{\bm x^h_{1m}, N {}^{(2)}C_0 + N^3 {}^{(2)} C_r + r^{-2}\delta N^e Z^e + \delta N^3 Z^h + \delta N Z^v\}\Big|_{\bm x^o = \bm x^h = \bm x^v=0, N = 1, N^3 = \sqrt{\frac{2M}{r}}}\\
    &=2 r \delta N^3_{1m} - \sqrt{2} \delta N^e_{1m} - \frac{1}{2} \pi_\mu^{(0)} \delta N_{1m} - \bm y_v^{1m}=0\\
    \dot{\bm x}^v_{1m} &= \{\bm x^v_{1m}, N {}^{(2)}C_0 + N^3 {}^{(2)} C_r + r^{-2}\delta N^e Z^e + \delta N^3 Z^h + \delta N Z^v\}\Big|_{\bm x^o = \bm x^h = \bm x^v=0, N = 1, N^3 = \sqrt{\frac{2M}{r}}}\\
    &=2 \partial_r \delta N^3_{1m} + \frac{\pi_\mu^{(0)}}{4 r^2} \delta N_{1m} + \frac{1}{r^2} \bm y_v^{1m} - \bm y_h^{1m} = 0
\end{align}
The differential equations are solved as follows: First, we solve the first equation for $\delta N^3_{1m}$ and sum the second with the third equation multiplied by $2 r^2$. This gives
\begin{align}
    \delta N^3_{1m} = - \frac{1}{\sqrt{2}}\qty(r^2 \partial_r (r^{-2} \delta N^e_{1m}) + \bm y_e^{1m})\\
    2 r \delta N^3_{1m} + 4 r^2 \partial_r \delta N^3_{1m} - \sqrt{2} \delta N^e_{1m} + \bm y_v^{1m} - 2 r^2 \bm y_h^{1m}=0
\end{align}
Inserting the first into the second we derive a differential equation for $\delta N^e_{1m}$:
\begin{align}
    - 2 \sqrt{2} r^2 \partial_r^2 \delta N^e_{1m} + 3 \sqrt{2} r \partial_r \delta N^e_{1m}  - 3 \sqrt{2} \delta N^e_{1m} - C_1^m - 4 \frac{C^m_2}{\sqrt{r}} =0
\end{align}
The solution of this differential equation is the sum of the general solution of the homogeneous equation and a particular solution of the inhomogeneous equation. The homogeneous solution grows with $r$ in the limit $r$ to infinity. Therefore, we need to choose the integration constants to be zero and the full solution is the particular solution which is given by
\begin{align}
    \delta N^e_{1m} = - \frac{C^m_1}{3 \sqrt{2}} - \frac{2 C^m_2}{3 \sqrt{2}\sqrt{r}}\,.
\end{align}
This determines the perturbed radial shift vector $\delta N^r$ and the perturbed lapse function $\delta N$. We have
\begin{align}
    \delta N^r_{1m} &= - \frac{C^m_1}{6 r} + \frac{C^m_2}{6\sqrt{r}^3}\,,\\
    \delta N_{1m} &= - \frac{2C_1^m}{\pi_\mu^{(0)}}\,.
\end{align}
In the work \cite{5} the authors demonstrate that these perturbations are related to the Schwarzschild geometry in an accelerated frame of reference.\\

To complete this section we insert our solutions for the dipole perturbations into the equation \eqref{eq:SolIntegralpimu2}. We obtain
\begin{align}
    4 \pi \sqrt{\frac{r_s}{r}}\pi_\mu^{(2)} &= \int \dd{r} \qty[\frac{1}{2} (\bm y_o)^2 + \frac{1}{2r^2}(\bm y_v)^2 + \frac{1}{2} (\bm y_e)^2 - \bm y_h \bm y_v] + \text{Terms with $l\geq2$}\\
    &= \int \dd{r} \sum_m  \qty[ \frac{a_m^2}{2 r^4} +\frac{1}{4}\qty(\frac{C^m_1}{r} + \frac{2 C^m_2}{\sqrt{r}^3})^2] + \text{Terms with $l\geq 2$}
\end{align}
The dipole contributions to $\pi_\mu^{(2)}$ vanish as $r$ approaches the boundary at infinity. Hence, the dipole perturbations will not be of interest for the final physical Hamiltonian.

\subsubsection{Odd Parity}

The reduced phase space for the odd-parity non-symmetric degrees of freedom is given by the quantities $\bm X^o, \bm Y_o$. The fall-off conditions of the true degrees of freedom is read off from the asymptotics in \eqref{eq:Decay}. We have that $\bm X^o$ grows at most linearly in $r$ and that $\bm Y_o$ vanishes like $r^{-2}$. In addition to $\bm X^o$, $\bm Y_o$, we have the gauge degrees of freedom $\bm x^o$ and $\bm y_o$. The degrees of freedom are related by one constraint $Z^o$. We will work in the gauge $\bm x^o = 0$ and solve the constraint for $\bm y_o$. We get
\begin{align}
    \bm y^{(1)}_o =  \frac{\sqrt{2(l+2)(l-1)}}{r^2} \int \dd{r} \qty( r^2 \bm Y_o + \frac{\pi_\lambda}{4r^2} \bm X^o)
\end{align}
The form of the constraint motivates a canonical transformation. The integral in the above solution for $\bm y_o$ is introduced as a new variable $\tilde Q$. Then, the conjugate variable $\tilde P$ is chosen such that these variables form a canonical pair. We define
\begin{align}
    \tilde Q &:= \sqrt{2} \int \dd{r} \qty(r^2 \bm Y_o + \frac{\pi_\lambda}{4 r^2} \bm X^o)\,,\\
    \tilde P &:= \frac{1}{\sqrt{2}}\qty(r^{-2} \bm X^o)'\,.
\end{align}
This transformation as well as the solution for $y_o^{(1)}$ is inserted into \eqref{eq:SolIntegralpimu2}. Then, the odd parity contributions with $l\geq2$ lead to 
\begin{align}
   4 \pi \sqrt{\frac{r_s}{r}}\pi_\mu^{(2)}\Big|_{l\geq 2, \text{odd}}= \int \dd{r} \frac{1}{4r}\pi_\mu^{(0)} \tilde P \cdot \tilde Q' + \frac{1}{2} \qty(r^2 \tilde P \cdot \tilde P + \frac{(l+2)(l-1)}{r^4}\tilde Q \cdot \tilde Q + \frac{1}{r^2} \tilde Q' \cdot \tilde Q')\,.
    \label{eq:OddParityTrafo1}
\end{align}
In the computation we neglected a boundary term which is given by
\begin{equation}
    - \int \dd{r} \dv{r}\qty(2 r^2 \tilde P \int \tilde P \dd{\tilde r} + \frac{1}{2}(2r + r_s) \qty(\int \tilde P \dd{\tilde r})^2)\,.
\end{equation}
Later, for the physical Hamiltonian, we are interested in $\pi_\mu^{(2)}$ in the limit as $r$ approaches infinity. From the orignal variables $\bm X^o$ and $\bm Y_o$ we determine that as $r$ approaches infinity $\tilde Q$ grows linearly and $\tilde P$ vanishes as $r^{-2}$. Therefore the boundary term vanishes like $r^{-1}$ and goes to zero in the limit.

It is possible to further simplify the contribution to $\pi_\mu^{(2)}$ using a canonical transformation from the variables $(\tilde Q, \tilde P)$ to new variables $(Q^o,P_o)$. We rescale the variable $\tilde Q$ by $r$ and shift $\tilde P$ to prevent any terms of the form $P_o Q^o$ form showing up. 
The transformation is given by
\begin{align}
    \tilde Q &= r Q^o\,,\\
    \tilde P &= \frac{1}{r} (P_o - \frac{\pi_\mu^{(0)}}{4 r^2} Q^o)\,.
\end{align}
We insert this transformation and after an integration by parts we have
\begin{align}
   4 \pi \sqrt{\frac{r_s}{r}}\pi_\mu^{(2)}\Big|_{l\geq 2, \mathrm{odd}}=\int \dd{r} &\frac{1}{4r}\pi_{\mu}^{(0)}P_o \cdot (Q^{o})' + \frac{1}{2} \Big(P_o \cdot P_o + Q^o{}' \cdot Q^o{}' + V_o Q^{o} \cdot Q^o \Big)\,,
\end{align}
with the Regge-Wheeler potential
\begin{equation}
    V_o =  \frac{1}{r^3} (l(l+1)r - 3 r_s )\,.
\end{equation}
The boundary term we neglected in the computation is  
\begin{align}
    \int \dd{r} \pdv{r}\qty(\frac{1}{2r^2}\qty(r - r_s) Q^o \cdot Q^o)\,.
\end{align}
The master variable $O^o$ is asymptotically constant and $P_o$ behaves as $r^{-1}$ for $r \to \infty$. The boundary term then vanishes as $r^{-1}$ and can be neglected. 

\subsubsection{Even Parity}

For the even parity case we do not follow the general strategy discussed in the rest of this series. Instead we solve the constraints for the variables $\bm x^h$, $\bm y_e$ and $\bm Y_e$. The steps are inspired by the methods that Moncrief used in his Hamiltonian analysis. The solution of the first order constraints is then inserted into the integral for $\pi_\mu^{(2)}$. We simplify the resulting expression using several canonical transformations and derive an expression that involves the master variables $Q^e, P_e$. In the end of this section we relate the master variables $Q^e$ and $P_e$ to the variables $\bm X^e$ and $\bm Y_e$ when we work in the GP gauge.

We start with the solution of the first order constraints. We proceed in two steps. First, we solve the diffeomorphism constraints ${}^{(1)}Z^h_{lm}$ and ${}^{(1)}Z^e_{lm}$ for the momenta $\bm y_e$ and $\bm Y_e$. Then, we apply a canonical transformation 
on the variables $(\bm x^v, \bm y^v)$ and $(\bm x^h,\bm y^h)$. This simplifies the first order Hamiltonian constraint $Z^v_{lm}$ so that it can be solved for $\bm x^h$.

We start by solving the constraint ${}^{(1)}Z^h_{lm}$ for $\bm y^e$ and the constraint ${}^{(1)}Z^e_{lm}$ for $\bm Y^e$. We use the gauge $\bm x^e = \bm X^e = 0$ and obtain
\begin{align}
    \bm y_e &= - \frac{1}{\sqrt{l(l+1)}}\Bigg(-2 \partial_r (\bm y_v) + 2 r \bm y_h - \partial_r( \pi_\mu^{(0)}) \bm x^v - \frac{1}{2} \pi_\mu^{(0)} \partial_r \bm x^v + \frac{\pi_\lambda}{2r^2}\partial_r \bm x^h\Bigg)\\
    \bm Y_e &=  - \frac{1}{r^2 \sqrt{2(l+2)(l-1)}}\Bigg(-\partial_r(r^2 \bm y_e) - \sqrt{l(l+1)}r^2 \bm y_h + \frac{1}{2} \sqrt{l(l+1)} \pi_\mu^{(0)} \bm x^v\Bigg)\,.
    \label{eq:SolDiffYe}
\end{align}
In the solution for $\bm Y_e$ we have to replace $\bm y_e$ with the solution for $\bm y_e$. 

We are now left with the degrees of freedom $(\bm x^v, \bm y_v)$ and $(\bm x^h, \bm y_h)$. Next, we consider the first order Hamiltonian constraint ${}^{(1)}Z^v_{lm}$. We insert the solution of the diffeomorphism constraint and implement the gauge $\bm x^e = \bm X^e = 0$. The result is 
\begin{align}
    {}^{(1)}Z^v_{lm} &=  \frac{1}{4 r^2} \pi_\mu^{(0)} \bm y_v- \frac{1}{2} \pi_\mu^{(0)} \bm y_h + 2 \qty(\partial_r^2 - \frac{1}{r}\partial_r -\frac{(l+2)(l-1)}{2 r^2}- \frac{r_s}{ r^3} ) \bm x^h -\qty(2 r \partial_r + l(l+1) + 2 - 2 \frac{r_s}{r}) \bm x^v\,.
\end{align}

Following Moncrief, we would like to solve this constraint for $\bm x^h$. There is an obstruction because the constraint depends on first and second derivatives of this variable. A possible solution to this problem is to introduce another canonical transformation of the variables to remove the derivative terms. We propose the following general form for the canonical transformation
\begin{align}
    \bm x^v &= q_1 + B q_2 + C \partial_r q_2 + D p_1\\
    \bm x^h &= q_2\\
    \bm y_v &= p_1 + G \partial_r q_2 \\
    \bm y_h &= p_2 - B p_1 + \partial_r \qty[(C - D G) p_1] - \partial_r (G q_1) + K q_2 - BG \partial_r q_2\,,
\end{align}
with arbitrary functions $B,C,D,G$ and $K$. We require that the first order Hamiltonian constraint is independent of the first and second derivatives of $q_2$. This imposes some conditions on the functions $C$ and $B$:
\begin{align}
    C =& \frac{1}{r}\,,\\
    B =& \frac{1}{(4 r^2 - r \pi_\mu^{(0)} G)}\qty[\frac{1}{2r}  \pi_\mu^{(0)} G + 4 \frac{r_s}{r} - 2 (l(l+1) + 2) ]\,.
\end{align}
Additionally we require the terms proportional to $p_1'$ to vanish. This leads to a condition on $D$:
\begin{equation}
    D = \frac{\pi_\mu^{(0)}}{r \pi_\mu^{(0)} G - 4 r^2}\,.
\end{equation}

The function $G$ is determined by looking at the solution of the diffeomorphism constraint for $\bm y^e$. It depends on up to two derivatives of $q_2$. We require the second derivative to vanish and obtain
\begin{align}
    G &= - \frac{\pi_\mu^{(0)}}{4r}.
\end{align}
The terms proportional to $q_2'$ vanish automatically by the background constraint equation ${}^{(0)}C_r=0$. The condition on $G$ gives a new form for B and D:
\begin{align}
    B =& \frac{1}{2 r (r + r_s)}\qty(\frac{r_s}{r} - (l(l+1)+2))\,,\\
    D = & - \frac{\pi_\mu^{(0)} }{4 r(r+r_s)}\,,
\end{align}

For the next steps in the analysis we used the symbolic computation features of Mathematica. The last function $K$ is determined by looking at the integral for $\pi_\mu^{(2)}$ in equation \eqref{eq:SolIntegralpimu2}. After applying the canonical transformation it depends on terms proportional $q_2 q_2''$, $q_2' q_2''$ and $(q_2')^2$. We can choose $K$ so that all of these terms vanish. Explicitly we have
\begin{align}
    K &= -\frac{\sqrt{r r_s} \left(\left(l (l+1) \left(l^2+l+4\right)-4\right) r^2+\left(l (l+1)
   \left(l^2+l+9\right)+4\right) r r_s+(3 l (l+1)+2) r_s^2\right)}{4 r^4 (r+r_s)^2}\,.
\end{align}

Using the full canonical transformation the first order Hamiltonian constraint is given by
\begin{align}
    \frac{l(l+1) \Lambda}{r^2} q_2 + \frac{\left(l^2+ l+2\right) r-3r_s}{r} q_1 + 
    2 (r+r_s) q_1'=0\,,
\end{align}
where $\Lambda = \frac{1}{2}(l+2)(l-1)+ \frac{3}{2} \frac{r_s}{r}$. In this form, the first order Hamiltonian constraint is solved for $q_2$. We obtain
\begin{align}
    q_2 = \frac{r^2}{l(l+1) \Lambda}\qty(\frac{\left(l^2+ l+2\right) r-3r_s}{r} q_1 + 
    2 (r+r_s) q_1')
\end{align}

Then, we apply the canonical transformation in the integral \eqref{eq:SolIntegralpimu2} and replace $q_2$ in terms of $q_1$ and $q_1'$. For $p_2$ we choose the gauge $p_2 = 0$. To further simplify the integral, we perform a final canonical transformation. We rescale the variables $q_1$ and $q_2$ by a factor and then shift $p_1$ in order that no cross terms of the form $Q^e P^e$ show up in the final result. 
\begin{align}
    q_1 =& \sqrt{\frac{l(l+1)}{(l+2)(l-1)}}\frac{\Lambda}{(r + r_s)} Q^e\\
    \begin{split}
    p_1 =& \sqrt{\frac{(l+2)(l-1)}{l(l+1)}} \frac{(r+r_s)}{\Lambda} \Bigg[ P_e + \frac{\sqrt{r r_s}}{{8 (l+2)(l-1) r^3 (r+r_s)^2 \Lambda}} \Big(6 \left(5 l^2+5 l-31\right) r_s^3\\
    &+(l+2)^2(l-1)^2 \left(l^4+2 l^3+13 l^2+12
   l+3\right) r^3+3 \left(9 l^4+18 l^3+5 l^2-4 l+35\right) r r_s^2\\
   &+2 (l+2)(l-1)(l^2 + l + 1) \left(5 l^2 + 5 l + 18\Big) r^2 r_s \right) Q^e\Bigg]
   \end{split}
\end{align}
We insert all the above transformations as well as the solutions for $\bm Y_e$, $\bm y_e$ and $q_2$ into Equation \eqref{eq:SolIntegralpimu2}. In the computation we use integration by parts to simplify the resulting expression significantly. The boundary term is recorded in appendix \ref{sec:EvenParityBoundaryTerm}. We will show now that it can be neglected. It is sufficient to only consider the leading contributions in the limit as $r$ approaches infinity. The others will then vanish automatically. The dominant boundary terms in $4\pi \sqrt{r/r_s} \pi_\mu^{(2)}$ are
\begin{align}
    &\qty(\frac{1}{2r} + O(r^{-2}))(q_2')^2 + \qty(- \frac{\left(l^2+l+2\right)}{r^2} + O(r^{-3})) q_2 q_2' +(2 + O(r^{-1}))q_2' q_1 - 3\frac{\sqrt{r r_s}}{r^2} q_2'p_1 \nonumber\\
    &+ \qty(-\frac{3(l^2+l+2)}{2 r}+ O(r^{-2})) q_2 q_1 + \qty(\frac{3 \sqrt{r r_s}(l^2+l+2)}{r^3} + O(r^{-7/2}))q_2 p_1\\
    &+ \qty(\frac{3 l^4+6 l^3+13 l^2+10 l+16}{8r^3} + O(r^{-4}))q_2^2 +\qty(\frac{3 r}{2} + \frac{(l^2 + l + 2)r}{l (l+1) (l+2)(l-1)} + O(1))q_1^2\nonumber\\
    &+ \qty(-\frac{1}{2r} + O(r^{-2})) p_1^2\nonumber+ \qty(- 6 \sqrt{\frac{r_s}{r}} + O(r^{-3/2})) q_1 p_1 + \qty(-\frac{1}{2r}+ O(r^{-2})) (Q^e)^2\nonumber
\end{align}

From the asymptotics in \eqref{eq:Decay} we read off the following boundary conditions on the canonical variables. The variable $\bm x^h$ grows linearly and $\bm x^v$ vanishes as $r^{-1}$. The conjugate momentum $\bm y_h$ vanishes like $r^{-2}$ and $\bm y_v$ behaves as a constant. From this we obtain that $q_1$ vanishes as $r^{-1}$ and $q_2$ grows linearly in $r$. The momentum $p_1$ behaves as a constant and $Q^e$ also as a constant. The full boundary term of $\pi_\mu^{(2)}$ vanishes as $r^{-1/2}$ and can be neglected.

The computation for the even parity contributions to $\pi_\mu^{(2)}$ lead to the integral
\begin{align}
    4 \pi \sqrt{\frac{r_s}{r}} \pi_\mu^{(2)}\Big|_{l\geq 2, \mathrm{odd}}=
    \int \dd{r} &\frac{1}{4}e^{-\lambda - 2 \mu} \pi_{\mu}^{(0)} P_e \cdot Q^e{}' + \frac{1}{2} e^{-2\mu + \lambda} \lambda'\Big(P_e \cdot P_e + Q^e{}' \cdot Q^e{}' + V Q^e \cdot Q^e\Big)\,,
\end{align}
where we introduced the Zerilli potential which has the form

\begin{equation}
    V_e = \frac{l(l+1) (l+2)^2(l-1)^2 + 3 (l+2)^2(l-1)^2 \frac{r_s}{r} + 9 (l+2)(l-1)\frac{r_s^2}{r^2} + 9 \frac{r_s^3}{r^3}}{4 r^2 \Lambda^2}\,.
\end{equation}

We end this section by showing that the same result for $\pi_\mu^{(2)}$ is obtained by 
working in the gauge $\bm x^v = \bm x^h = \bm x^e=0$ and solving for $\bm y_v,\bm y_h,\bm y_e$ keeping $\bm X^e,\bm Y_e$ as the true degrees of freedom. For the discussion we need some aspects of the reduced phase space quantization program that we recall in the following.

We consider a system with constraints $C(x,y,X,Y)$ which as in our 
calculations depend on two sets of canonical variables $x,y$ and $X,Y$. 
In the following paragraphs we will not display the symmetric degrees 
of freedom as they are not important here. 
According to the reduced phase space quantisation program there are now two options. Either we consider $x,y$ as the true and $X,Y$ as the gauge degrees of freedom or the other way around. Hence, we  solve the constraint equation $C=0$ as $Z = Y + H(X,x,y)$ or as $z = y + h(x,X,Y)$. In the first case we will work in the gauge $G = X = 0$ and in the second case we will consider the gauge $g = x = 0$.

Associated to each of the two presentations $z$ and $Z$ there is a projector onto gauge invariant functions that we denote by $O$ and $o$. These maps are defined for any phase space function $F$ as
\begin{align}
    O_F&:=[e^{V_{Z(S)}}\cdot F]_{S=-X}\\
    o_F&:=[e^{V_{z(s)}}\cdot F]_{s=-x}\,,
\end{align}
where $V_{A}$ is the Hamiltonian vector field associated to the phase 
space function $A$. In these equations $s,S$ are smearing fields. Given a general function $F(x,y,X,Y)$ on phase space we can determine the action of $O,o$ explicitly. We have $[O_F]_{X=0}=F(x,y,0,-H(0,x,y))$ and $[o_F]_{x=0}=F(0,-h(0,X,Y),X,Y)$ where we also implemented the appropriate gauge fixings $g,G$. 

In \cite{I} it is shown
that the physical Hamiltonian corresponding to the gauge choice 
$G$ can be described as the restriction to $X=0, Y=-H(X=0,x,y)$ of a  
boundary functional $B(x,y,X,Y)$.
We can now study it in the two gauges $G,g$. 
Using the maps $O,o$ we define $E_{X=0} = [O_B]_{X=0} = B(x,y,0,-H(x,y,0)$ and $e =[o_B]_{x=0} = B(0,-h(0,X,Y),X,Y)$. In general the relation between $E$ and $e$ is not immediately clear. Let us therefore not implement the gauge fixing conditions $G,g$. Then, in general $E,e$ are functions of $X,x$ respectively and we write $E = E(O_x, O_y,X)$ and $e = e(o_X,o_Y,x)$. The explicit dependence of $E,e$ on $X,x$ is reflected the fact that the function $B$ 
is in general not gauge invariant. 
In fact, if the function $E,e$ were independent of $X,x$ 
respectively, the function $B$ would be weakly gauge invariant. 
Notice that for non gauge invariant $B$, there would be no way to 
recover the gauge variant contributions from the gauge fixed versions 
$E_{X=0}$ or $e_{x=0}$. We see that a relation between $O_B,o_B$ 
can only be established if $B$ is weakly gauge invariant. Note that 
a possible difference between $o_B, O_B$ is not contradictory, 
this would just reflect the fact that a Hamiltonian depends on the 
frame of the observer. 

Note that 
weak invariance of $B$ would be automatic if $B$ was just the boundary 
term $b$ that one needs to add to $C$ in order that the smeared constraint
$H(f)=C(f)+b(f)$ was differentiable. This is because the $H(f)$ close among 
themselves \cite{Thiemann2007}
for phase space independent smearing functions $f$, thus if 
$f$ corresponds to a gauge transformation (i.e. $H(f)=C(f)$) and 
and $g$ a symmtery transformation (i.e. $H(g)\not= C(g)$)
then $\{H(f),H(g)\}=\{C(f),H(g)\}=H([f,g])=C([f,g])$ because the bracket 
$[f,g]$ corresponds to a gauge transformation when one of $f,g$ is a 
gauge transformation. This means that $H(g)$ is a weak Dirac observable
for any phase space independent smearing function $g$ 
which weakly coincides with $b(g)$. However, as shown in \cite{I},
our $B$ is more complicated: The physical Hamiltonian 
is of the form $H=b_\ast(g_\ast)/2$ where 
$_\ast$ instructs to evaluate at the solution of the constraints, gauge 
conditions $G$ and stability conditions. It can also be written as 
$\chi(j)_{j=j_\ast}$ where $j$ is a function on the full phase space
and $j_\ast$ its restriction to $G=0=C$. Thus while the boundary 
term $B=\chi(j)$ is an extension 
of the physical Hamiltonian to the full phase space its 
weak invariance is not at all obvious due to the fact that $G=0$ was used 
to obtain the extension $B$. 

Assume that nevertheless
we can show that $B$ is weakly gauge invariant, at least 
to second order. Then, we have that $E \approx e$ and we compute
\begin{align}
    \begin{split}
    E(x,y) &= E(O_x,O_y)_{X=0} = [O_B]_{X=0}=[{O_{O_B}}]_{X=0}\\
    &= [O_{o_B}]_{X=0} = [O_{e}]_{X=0} = e([O_{o_X}]_{X=0},[O_{o_Y}]_{X=0})\\
    &=e\qty([o_X]_{X=0,Y=-H},[o_Y]_{X=0,Y=-H})\,.
    \end{split}
\end{align}
In the computation, we used the fact that on the gauge cut $G=X=0$, $O_x,O_y$ reduce to $x,y$ respectively. In the third step we used that $O_F$ is a projector and in the fourth step, the weak gauge invariance of $B$. For the last step we put the action of the projector $O$ explicitly. The reverse direction expressing $e$ in terms of $E$ works similarly. This shows that for weakly gauge invariant $B$ we can work in any of the two gauges because we can always relate the two results.

We now adapt the above arguments to our situation. The analysis consists of 
two steps. First, we have to establish that the physical Hamiltonian 
constructed from $\pi_\mu^{(2)}$ is weakly gauge invariant to 
second order.
Then, in the second step, we can express the variables $q_1, p_1$ in terms of the degrees of freedom $\bm X^e,\bm Y_e$.

The physical Hamiltonian, as we will see, depends on $\pi_\mu^{(2)}$ only in the limit as $r$ tends to infinity. Therefore, it is sufficient to show that the gauge-variant contributions to $\pi_\mu^{(2)}$ are vanishing in this limit. The formula \eqref{eq:SolIntegralpimu2} for $\pi_\mu^{(2)}$ involves an integral over the symmetric second order constraints. If the only contributions which are not gauge invariant was in a boundary term which vanishes at infinity the physical Hamiltonian would be gauge invariant. In the following we will show that this is the case for the even parity sector.

In the calculations of this section we performed a canonical transformation of the variables $\bm x^h, \bm y_h,\bm x^v, \bm y_v$ to new variables $q_1,p_1$ and $q_2, p_2$. In these degrees of freedom we solved the constraints for $\bm Y_e, \bm y_e, q_2$. In the following we need the solutions in a slightly more general form where we keep the degrees of freedom $\bm X^e$, $\bm x^e$ and $p_2$. In this setup the solutions of the first-order even parity constraints are
\begin{align}
\begin{split}
    q_2^{(1)} &= \frac{r^2}{l (l+1) \Lambda}\Bigg(\sqrt{\frac{(l-1) l (l+1) (l+2)}{2}} \frac{\bm X^e}{r^2} - \frac{2}{r} \sqrt{l(l+1)} \partial_r (r \bm x^e{}) + \qty((l^2+l+2) - 3 \frac{r_s}{r}) q_1\\
    &~~~~~~~~~~~~~~~~~~+2 \qty(r + r_s) q_1' + 2 p_2 \sqrt{r r_s}\Bigg)
\end{split}\\
    \begin{split}
        \bm y_e^{(1)} &= \frac{3 \sqrt{r r_s}}{\sqrt{l(l+1)} r} q_1 - \frac{\sqrt{l(l+1)}}{r+r_s} p_1+ \frac{\sqrt{l(l+1)}\sqrt{r r_s} \qty((l^2 + l + 6)r + 3 r_s) }{2 r^3( r+ r_s)} q^{(1)}_2\\
        &- \frac{2r}{\sqrt{l(l+1)}}p_2 - \frac{\sqrt{r r_s}}{r^2}\bm x^e
    \end{split}\\
        \bm Y_e^{(1)} &= -\frac{\sqrt{r r_s} \qty((l^2+l+5) r + 2 r_s)}{4 r^4 (r+ r_s)} \bm X^e + \frac{\sqrt{r r_s} \qty((l^2+l+6) r+3 r_s)}{\sqrt{2(l+2)(l-1)} r^2 (r+ r_s)}\bm x^e{}' + \frac{\sqrt{r r_s} \qty((2 l (l+1)+9) r+3 r_s)}{2 \sqrt{2 (l+2)(l-1)}r^3 (r+ r_s)}\bm x^e \nonumber\\
        &+ \sqrt{r r_s}\frac{\qty(3 -l(l+1) (l^2+l+10) )r - 2 (l^2+l-9) r_s}{2 \sqrt{2 (l-1) l (l+1) (l+2)} r^2 (r+r_s)} q_1 + \sqrt{\frac{l(l+1)}{2(l+2)(l-1)}}\frac{\Lambda}{r (r+r_s)} p_1\\
        &+ \frac{\left(l^2+l-6\right) r-9 r_s}{\sqrt{2(l-1) l (l+1) (l+2)}(r+r_s)} p_2 -\frac{2 r}{\sqrt{2(l-1) l (l+1) (l+2)}}p_2'\nonumber
\end{align}

The solutions of the constraints define the projector $O$ onto the gauge invariant phase space functions. All the constraints are linear in the perturbations, therefore, it is sufficient to terminate the infinite series after the first order. We have for any phase space function $F$ that 
\begin{align}
    O_F = F + \int \dd{\tilde r}\qty[ p_2(\tilde r)\{q_2(\tilde r) - q_2^{(1)}(\tilde r),F\} + \bm x^e(\tilde r)\{\bm y_e(\tilde r) - \bm y_e^{(1)}(\tilde r),F(r)\} + \bm X^e(\tilde r)\{\bm Y_e(\tilde r) - \bm Y_e^{(1)}(\tilde r),F\}]\,.
\end{align}

The remaining degrees of freedom $q_1, p_1$ are the true degrees of freedom. They are extended to gauge invariant variables using the map $O$ defined above. This gives the following functions:
\begin{align}
    O_{q_1} &= q_1 - \frac{\sqrt{l(l+1)}}{(r+r_s)} \bm x^e + \sqrt{\frac{l(l+1)}{2 (l+2)(l-1)}}\frac{\Lambda}{r^2 (r+r_s)}\bm X^e\\
    O_{p_1}&= p_1 - \sqrt{r r_s}\frac{\qty(3 -l(l+1) (l^2+l+10) )r - 2 (l^2+l-9) r_s}{2 \sqrt{2 (l-1) l (l+1) (l+2)} r^2 (r+r_s)} \bm X^e + \partial_r\qty(\frac{\sqrt{r r_s} \qty((l^2 + l + 6)r + 3 r_s) }{\sqrt{l(l+1)} r\Lambda} \bm x^e)\nonumber\\
    &-\qty(\frac{\sqrt{r r_s} \qty((l^2 + l + 6)r + 3 r_s) \qty((l^2+l+2)r - 3 r_s) }{2\sqrt{l(l+1)} r^2( r+ r_s)\Lambda}+ \frac{3 \sqrt{r r_s}}{\sqrt{l(l+1)} r}) \bm x^e\\
    &-\partial_r\qty(\frac{2 r^2 (r+r_s)}{l (l+1) \Lambda}p_2) - \frac{r^3 \qty((l^2 + l + 2) - 3 \frac{r_s}{r})}{l (l+1) \Lambda}p_2\nonumber
\end{align}
By construction $O_{q_1}, O_{p_1}$ are gauge invariant and commute with the constraints $q_2 - q_2^{(1)}$, $\bm y_e - \bm y_e^{(1)}$ and $\bm Y_e - \bm Y_e^{(1)}$.

To show the weak gauge invariance of $\pi_\mu^{(2)}$ we replace $q_1$ and $p_1$ by their gauge invariant extensions $O_{q_1},O_{p_1}$ respectively. Then, using Mathematica we can show that the terms involving 
derivatives of $q_2$ is a boundary term. It is given in appendix \ref{sec:BoundaryTermProofGaugeInv} as the quantity $A_1$. Next, we remove $q_2$, $\bm y_e$ and $\bm Y_e$ using the solution of the constraints. A symbolic calculation with Mathematica shows that all of the terms involving the gauge degrees of freedom $\bm x^e$, $\bm X^e$ and $p_2$ are equal to a boundary term proportional to $A_2 + A_3 + A_4$ where the $A_i$ are defined in appendix \ref{sec:BoundaryTermProofGaugeInv}. Therefore, we showed that
\begin{align}
    \pi_\mu^{(2)} = \frac{1}{4\pi}\sqrt{\frac{r}{r_s}}\int I(O_{q_1},O_{p_1})\dd{r} +\frac{1}{4\pi}\sqrt{\frac{r}{r_s}}\qty(A_1 + A_2 + A_3 + A_4)\,,
\end{align}
where $I(O_{q_1},O_{p_1})$ is gauge invariant and the gauge dependent terms are all in the $A_i$. If we show that the sum of the $A_i$ vanishes in the limit $r$ to infinity the remaining integral is expressed fully in terms of gauge invariant variables and $\pi_\mu^{(2)}$ is weakly gauge invariant.

We now provide the leading order in $r$ contribution to the boundary term. We use the following convention for the asymptotic behaviour of the canonical variables which are based on equation \eqref{eq:Decay}:
\begin{align}
    q_1 &\sim q_1^0 r^{-1}, & p_1 &\sim p_1^0, &O_{q_1} &\sim \overline q_1^0 r^{-1}, & O_{p_1} &\sim \overline p_1^0, & q_2 &\sim q_2^0 r, & p_2 &\sim p_2^0 r^{-2},\nonumber\\
    \bm X^e&\sim \bm X^e_0 r, & \bm x^e&\sim \bm x^e_0\,.
\end{align}
The quantities $q_1^0,p_1^0,\overline q_1^0,\overline p_1^0,q_2^0,p_2^0,\bm X^e_0,\bm x^e_0$ are constant with respect to the radius $r$ but are in general still dependent on $l$ and $m$. We have
\begin{align}
    \sum_{i=1}^4 A_i &=\frac{1}{r}\Big[\frac{3 l^4+6 l^3-5 l^2-8 l+8}{16} (\bm X^e_0)^2 -\frac{3}{4} \sqrt{2 (l-1)l(l+1)(l+2)} \bm X^e_0 \overline q^1_0 + \frac{8 p_2^0 \left(l(l+1) \overline p_1^0 - 2 p_2^0\right)}{(l-1) l^2 (l+1)^2 (l+2)}\nonumber\\
    &+ \frac{(l-1) l (l+1) (l+2)}{\sqrt{2(l+2)(l-1)}} \bm X^e_0 \bm x^e_0 - \frac{1}{2} \left(l^2+l+4\right) (\bm x^e_0)^2 - 2\sqrt{l (l+1)} \bm x^e_0 \overline q_1^0 - \frac{2 \left(l^2+l-4\right) \overline p_1^0 p_2^0}{l(l+1) \left(l^2+l-2\right)}\nonumber\\
    &+q_2^0 \left(\frac{1}{4} \sqrt{2(l-1) l (l+1) (l+2)} \bm X^e_0  - \frac{1}{2}\sqrt{l(l+1)}^3 \bm x^e_0 - \frac{1}{8} (3 l(l+1)+2) (l(l+1) q_2^0 - 4 q_1^0)-\frac{1}{2} q_2^0\right)\nonumber\\
    &-\frac{4 \left(3 l^2+3 l-8\right) (p_2^0)^2}{(l-1)^2l^2
   (l+1)^2(l+2)^2}\Big] + O(r^{-3/2})
\end{align}
We see that the leading contributions vanish like $r^{-1}$ as $r$ goes towards infinity. This shows that in this limit the boundary term vanishes and that $\pi_\mu^{(2)}$ is weakly gauge invariant. Therefore, it is justified to simply relate the variables between the different gauge fixings.

We then move on to relate the variables in the two different gauge fixings. The above calculations already provide some of the necessary formulas. In the main analysis we worked in the gauge $\bm x^e = \bm X^e = p_2 = 0$. We would like to change this to the gauge $\bm x^e=\bm x^h=\bm x^v=0$. Therefore, we take the expression for $O_{q_1},O_{p_1}$ and perform the necessary substitutions. For $O_{q_1}$ we have
\begin{align}
    O_{q_1} = D \bm y_v + \sqrt{\frac{l(l+1)}{2 (l+2)(l-1)}}\frac{\Lambda}{r^2 (r+r_s)}\bm X^e \,,
\end{align}
where we replaced $q_1$ by $D \bm y_v$ and used the gauge condition $\bm x^e = 0$. The relation between $q_1$ and $D \bm y_v$ follows from reversing the first canonical transformation and using the gauge fixing $\bm x^h= \bm x^v=0$. The momentum $O_{p_1}$ is given by
\begin{align}
\begin{split}
    O_{p_1} &= \bm y_v - \sqrt{r r_s}\frac{\qty(3 -l(l+1) (l^2+l+10) )r - 2 (l^2+l-9) r_s}{2 \sqrt{2 (l-1) l (l+1) (l+2)} r^2 (r+r_s)} \bm X^e\\
    &-\partial_r\qty(\frac{2 r^2 (r+r_s)}{l (l+1) \Lambda}p_2) - \frac{r^3 \qty((l^2 + l + 2) - 3 \frac{r_s}{r})}{l (l+1) \Lambda}p_2
\end{split}
\end{align}
where
\begin{align}
    p_2 &= \bm y_h + B \bm y_v + \partial_r (C \bm y_v)\,.
\end{align}

We conclude the discussion by providing the solutions of $\bm y_v$ and $\bm y_h$ in terms of $\bm X^e, \bm Y_e$. This is achieved by solving the first order constraints for $\bm y_v$ while keeping $\bm X^e$ and $\bm Y_e$ free. The constraint equation $Z^v_{lm}=0$ gives
\begin{align}
    \bm y_h = \frac{\bm y_v}{2r^2} - \frac{1}{r^2 \pi_\mu^{(0)}} \sqrt{2(l+2)(l+1)l(l-1)} \bm X^e\,.
    \label{eq:Solyhyv}
\end{align}
We use this in $Z^h_{lm}=0$ to obtain an expression for $\bm y_e$:
\begin{align}
    \sqrt{l(l+1)} \bm y_e = 2 \partial_r \bm y_v - \frac{1}{r} \bm y_v + \frac{2}{r\pi_\mu^{(0)}}\sqrt{2 (l+2)(l+1)l(l-1)} \bm X^e\,.
\end{align}
Then, $Z^e_{lm}=0$ reduces to a differential equation of the form:
\begin{align}
    - 2 r^2 \partial_r^2 \bm y_v - 3 r \partial_r \bm y_v - \frac{(l+2)(l-1)}{2} \bm y_v = s(r)\,,
\end{align}
with a ``source'' term $s(r)$ depending on $\bm X^e, \bm Y_e$
\begin{equation}
    s(r):=- \sqrt{2(l+2)(l+1)l(l-1)}\qty(r^2 \bm Y^e + \frac{\pi_\lambda}{4 r^2} \bm X^e + \frac{l(l+1) - 1}{\pi_\mu^{(0)}} \bm X^e - \frac{2r}{\pi_\mu^{(0)}} \partial_r \bm X^e)\,.
\end{equation}
The differential equation for $\bm y_v$ is a linear, inhomogeneous second order differential equation. The solution is the sum of the general solution to the homogeneous equation and a particular solution of the inhomogeneous equation. The homogeneous solution is
\begin{align}
    \bm y_v = C_+ r^{\alpha_+} + C_- r^{\alpha_-}\,,
\end{align}
with two integration constants $C_{\pm}$ and the parameter $\alpha_\pm$ defined as
\begin{equation}
    \alpha_\pm = - \frac{1}{4} \pm \frac{i}{4}\sqrt{4(l+2)(l-1) -1}\,.
\end{equation}
A particular solution is given by
\begin{align}
    \bm y_v = r^{\alpha_-} \int  \dd{\tilde r}\qty(\frac{2 i \tilde r^{\alpha_+ - \frac{1}{2}}}{\sqrt{4(l+2)(l-1)-1}} s(\tilde r)) - r^{\alpha_+} \int \dd{\tilde r} \qty(\frac{2 i \tilde r^{\alpha_- - \frac{1}{2}}}{\sqrt{4(l+2)(l-1)-1}} s(\tilde r)) 
\end{align}
Therefore we successfully related $\bm y_v$ to the true degrees of freedom $\bm X^e, \bm Y_e$. The relation for $\bm y_h$ then follows from equation \eqref{eq:Solyhyv}. This completes the relation of $O_{q_1}$ and $O_{p_1}$ to the variables $\bm X^e$ and $\bm Y_e$ and shows that working in our gauge is completely equivalent to working in the GP gauge proposed in the rest of this series of papers.

\subsection{The physical Hamiltonian}

As already mentioned, the reduced Hamiltonian is a boundary term which is derived 
in \cite{I}. To make this paper self-contained we will describe briefly how this 
works.

One needs to impose the decay behavior of the canonical variables in the limit as $r$ 
tends towards infinity. This is derived from the ususal 
decay behaviour by translating it into Gullstrand-Painlevé coordinates. 
We showed that in these coordinates $\pi_\mu$ and $\pi_\lambda$ behave 
as $O^{\infty}(\sqrt{r})$ at infinity. 
The symplectic term of the action is well defined provided 
that the conjugate variables vanish according to $\mu=O^{\infty}(r^{-\frac{3}{2}-\epsilon})$ 
and $\lambda=\log(r) + O^{\infty}(r^{-\frac{3}{2}-\epsilon})$ for 
$\epsilon>0$.
This is certainly satisfied for the Gullstrand-Painlevé gauge conditions ($\mu=0$ and $\lambda = \log(r)$).  

We also require the lapse function and shift vector to have appropriate decay behaviours in order that the 
symplectic structure and constraints have finite values and 
are functionally differentiable including the 
boundary terms. This in particular turns out to impose that
$N^3=S^3 = O^\infty(r^{-1/2})$ and $N=S^0 =O^\infty(1)$. The value of lapse and shift 
are not arbitrary but are dictated by the stability condition of the GPG under 
the transformations generated by the constraints including the boundary term which 
turns them into explicit expressions $S_\ast$ in terms of the true degrees of freedom $X,Y$
and all decay conditions must work consistently together so that $S_\ast$ falls into 
the allowed class of functions. 
Thus eventually $x=x_\ast=0, y=y^\ast, q=q_\ast, p=p^\ast, S=S_\ast$ are concrete 
functions of $X,Y$.

The physical or reduced Hamiltonian is that function $H$ of $X,Y$ which 
has the same Poisson brackets with any function $F$ of $X,Y$ as the function
$C(f)+Z(g)+B(f,g)$ followed by evaluation at 
$x=x_\ast=0, y=y^\ast, q=q_\ast, p=p^\ast, f=S_\ast, g=g\ast$ where $f$ is the 
$l=0$ mode of $S$ and $S=f+g$ and $B(f,g)$ is the boundary term. The analysis 
of \cite{I} yields the non-perturbative but implicit result
\begin{align}
H&= \lim_{r \to \infty} \frac{c}{\kappa r}\int_{S_r} \frac{(W^{33})^2}{\sqrt{\det \Omega}}  \dd \Omega\\
&= \lim_{r\to \infty} \frac{\pi c}{\kappa r} \pi_\mu^2
\end{align}
where $\kappa=16\pi$ is the gravitational coupling constant in units $G=1$ and $c$ is an integration constant. It can be interpreted as the ticking 
rate of the GP time clock. If one wants the lapse to asymptote to unity then
$c=1/2$.
In perturbation theory one is instructed to expand $H$ to any desired order 
relying on the perturbative construction of $\pi_\mu$.     
  
To second order we find for $c=1/2$
\begin{align}
    \lim_{r\to \infty} \frac{\pi}{2 \kappa r} \pi_\mu^{(0)}
    \qty(\pi_\mu^{(0)} + 2\;\pi_\mu^{(2)})
    = M + \frac{1}{\kappa} \int_{\mathbb{R}^+} \dd{r} \sqrt{\frac{r_s}{r}} {}^{(2)}C_h + {}^{(2)}C_v\,.
\end{align}

The fact that $\pi_\mu^{(2)}$ enters only linearly despite the fact that 
the full expression involves a square is responsible for the 
fact that even and odd contributions decouple.
Inserting the concrete expressions found we obtain the reduced Hamiltonian
\begin{equation}
    H = M + \frac{1}{\kappa} \sum_{l\geq 2,m,I \in \{e,o\}} \int_{\mathbb{R}^+} 
\dd{r} [N^3 P^I_{lm} \partial_r Q^I_{lm} + \frac{N}{2} \;((P^I_{lm})^2 + (\partial_r Q^I_{lm})^2 + V_I (Q^I_{lm})^2)]
\end{equation}
where we restored the labels $l$ and $m$ and $Q,P$ are the functionals of $\bm X, \bm Y$ displayed 
in the above subsections. We also have introduced the zeroth order lapse $N=1$ and shift 
$N^3=\sqrt{2M/r}$ in the GPG. In the Hamiltonian, the potential $V_o$ is the Regge-Wheeler and $V_e$ the Zerilli potential. 

The method we used to derive the physical Hamiltonian to second order also incorporates the zeroth order background contribution.
The advantage of this is that the relative normalisation between the zeroth and second order terms is determined. 
To our knowledge this goes beyond what has been studied in the literature so far, where only the second order contributions to the Hamiltonian have been considered. 
For instance, the previous treatments in \cite{6,7,8} do not take boundary terms into account. While boundary terms are not essential for achieving their results, a more careful treatment has to include them. As pointed out for example in \cite{Benguria1977}, boundary terms are a necessary ingredient when studying spherically symmetric spacetimes. For the Schwarzschild solution one finds that the boundary is given by the Schwarzschild mass consistent with the above physical Hamiltonian. Neglecting this term we would find no zeroth order contribution since by definition the background constraints vanish for the Schwarzschild solution. 

As in the previous works we had to integrate by parts several times in the derivation of the physical Hamiltonian. This lead to boundary terms which we proved to vanish under the fall-off conditions for the canonical variables. Before, to the best of our knowledge such boundary terms were neglected in the literature without further justification. Our analysis provides the missing details and shows why we are allowed to drop these terms.

Thus our alternative method to compute the reduced Hamiltonian driving the dynamics
of the gauge invariant perturbations has reproduced the known results to second 
order. The real virtue of our method is of course that it immediately 
generalises to any higher order leading to self interacting graviton $X^3$,$X^2 Y$, $X Y^2$, $Y^3$,$\dots$ 
contributions from third order onwards.

\section{Relation to Lagrangian Formalism:}
\label{sec:Comparison}

For a check of consistency we compare our result in the Hamiltonian framework with the Lagrangian formulation 
reviewed in Appendix \ref{sec:PerturbLagrangian}. For this we study the Hamiltonian equations of motion. The contributions to the physical Hamiltonian were all brought to a similar form which differ only by the potential term. The general form is given by
\begin{align}
    \frac{1}{\kappa}\int \dd{r} N^3 P Q' + \frac{N}{2}\qty(P^2 + (Q')^2 + V Q^2)\,.
\end{align}
For the following investigations we set $\kappa=1$. The Hamiltonian equations of motion for the above Hamiltonian are
\begin{align}
    \dot Q &= N^3 Q' + N P\,,\\
    \dot P &= \partial_r (N^3 P) + \partial_r (N Q') - N V Q\,.
\end{align}
The first equation is solved for $P$ and inserted into the second equation to give
\begin{align}
    - \partial_t \qty(\frac{1}{N}\qty(\dot Q - N^3 Q')) + \partial_r \qty(\frac{N^3}{N}\qty(\dot Q - N^3 Q')) + \partial_r (N Q') - N V Q =0\,.
\end{align}
Notice that the metric on the $(t,r)$ slice is given by
\begin{align}
    g = \mqty(-N^2 + (N^3)^2 &  N^3\\  N^3 & 1) ~~~~~~~~ g^{-1} = \mqty(-\frac{1}{N^2} & \frac{N^3}{N^2}\\ \frac{N^3}{N^2} & 1 - \frac{(N^3)^2}{N^2})\,
\end{align}
with $\sqrt{-\det(g)}= N$. Therefore dividing the equation by $N$ and rearranging the terms we get
\begin{align}
    \frac{1}{\sqrt{-g}}\partial_t\qty(\sqrt{-g}\qty(g^{tt} \dot Q + g^{tr} Q')) + \frac{1}{\sqrt{-g}}\partial_r \qty(\sqrt{-g}\qty(g^{rt} \dot Q + g^{rr} Q')) = V Q\,.
\end{align}

The differential operator on the left hand side is just the Laplace-Beltrami operator associated to $g$. Therefore, we obtain the wave equation
\begin{equation}
    \square Q = V Q\,.
\end{equation}
We now simply compare the potentials and see that they match. In the odd parity sector we obtain the Regge-Wheeler equation for $Q^o$ and for the even parity variables we get the Zerilli equation for $Q^e$. 

To complete the comparison of the Lagrangian and the Hamiltonian formulation we relate the master variables. We need to show that the quantities $Q^{o/e}$ defined in the Hamiltonian setup agree up to some constant factor with the master variable $\psi$ introduced in the appendices. Let us start with the odd parity variables. In the Hamiltonian calculation we showed that
\begin{equation}
    Q^o = \frac{\sqrt{2}}{r}\int \dd{r} \qty(r^2 \bm Y_o + \frac{\pi_\lambda}{4r^2} \bm X^o)\,,
\end{equation}
satisfies the Regge-Wheeler wave equation.

In Appendix \ref{sec:PerturbLagrangian} we derived the same wave equation form linearized gravity. The master variable was defined as
\begin{align}
    \psi =  r^3 \epsilon^{t3} \qty(\partial_t( r^{-2} \tilde h_r) - \partial_r(r^{-2} \tilde h_t))\,,
\end{align}
where $\tilde h_t$ and $\tilde h_r$ are gauge invariant variables. Their relation to the variables $h_a$ and $h^o$ is given in the appendices. We adapt the notation to the Hamiltonian formalism used in the main text of the manuscript. We have 
\begin{align}
    \tilde h_r &= \bm x^o - \frac{r^2}{\sqrt{2(l+2)(l-1)}} \partial_r (r^{-2}\bm X^e)\\
    \tilde h_t &= h^o_t - \frac{1}{\sqrt{2(l+2)(l-1)}} \partial_t \bm X^e
\end{align}
In the Hamiltonian theory we have the perturbed shift vector. For the odd parity sector the radial component vanishes, $\delta N^3 = 0$. The angular components are $\delta N^A = \sum_{l,m} r^{-2} h_{t,lm}^o L_{o,lm}^A$. We insert the expressions for $\tilde h_r$ and $\tilde h_t$ into the master variable $\psi$. Then, we observe that the terms with $\bm X^e$ cancel. We are left with
\begin{align}
    \psi = - r^3 \epsilon^{t3} \partial_r (r^{-2} h^o_t)
\end{align}
The shift vector is determined using the stability condition of the gauge fixing $\bm x^o=0$. We obtain
\begin{align}
    \dot {\bm x^o} &= \qty{\bm x^o,N {}^{(2)}C_v + N^3 {}^{(2)} C_h + r^{-2} h_t \cdot Z^o}_\ast\\
    &= \bm y_o^{(1)} + r^2 \partial_r (r^{-2} h_t) \,.
\end{align}
The result of the Poisson bracket is evaluated on the respective gauge cut $\bm x^o=0$, $\bm y_o = \bm y_o^{(1)}$. For the lapse function and shift vector we used the expressions in GP gauge, i.e. $N=1$ and $N^3 = \sqrt{\frac{r_s}{r}}$. The master variable is then given by
\begin{align}
    \psi = r \epsilon^{t3} \bm y^{(1)}_o\,.
\end{align}
The final step is to insert the solution $\bm y^{(1)}_o$ of the first order diffeomorphism constraint. We obtain
\begin{align}
    \psi = \epsilon^{t3} \frac{\sqrt{2(l+2)(l-1)}}{r} \int \dd{r} \qty( r^2 \bm Y_o + \frac{\pi_\lambda}{4r^2} \bm X^o)
\end{align}
Finally, we use the definition of the Levi-Civita pseudotensor ($\epsilon^{t3}=1$). The odd parity master variable of the Hamiltonian and Lagrangian approach coincide up to an $l$-dependent prefactor.

We now move to the even parity master variables. In the main text we constructed the variable $Q^e$ in the Hamiltonian formulation and in Appendix \ref{sec:PerturbLagrangian} we studied the master variable $\psi_Z$. Both of them satisfy the Zerilli wave equation. The variable $\psi_Z$ is defined as
\begin{equation}
    \psi_Z = \frac{1}{l(l+1)}\qty(\gamma K + \frac{1}{\Lambda}\qty(\gamma \gamma^a \gamma^b k_{ab} - \gamma^2 \gamma^a \nabla_a K))\,.
\end{equation}
Similarly to the odd parity case we replace the variables with corresponding ones in the Hamiltonian formulation. In the appendix we had the following relations for $K$ and $k_{ab}$:
\begin{align}
    K &= r^{-2} \bm x^h - \frac{2}{\sqrt{l(l+1)}r}  \gamma^a j_a\,,\\
    k_{ab} &= h_{ab} - \frac{1}{\sqrt{l(l+1)}}\qty(\nabla_a j_b + \nabla_b j_a)\,.
\end{align}
By our choice of gauge for the perturbations we set $G$ and $j_r$ equal to zero. The variables $h_{tt}$, $h_{tr}$ and $j_t$ are related to perturbations of lapse function and shift vector. We take $N + \delta N$ for the lapse function and $N^3 + \delta N^3$ and $\delta N^e Y_e^A$ for the shift vector. Then we have  
\begin{align}
    h_{tt} &= - 2 N \delta N + \bm x^v (N^3)^2 + 2 N^3 \delta N^3\,,\\
    h_{t3} &= \delta N^3 +  \bm x^v N^3\,,\\
    h_{33} &= \bm x^v\,,\\
    j_t &= r^2 \delta N^e\,,\\
    j_3 &=0\,.
\end{align}

Next, we calculate the components of $\gamma^a$. By definition we have $\gamma_a = \partial_a(r)$. This gives $\gamma_t = 0$ and $\gamma_3 = 1$. Raising the index with the inverse metric we obtain
\begin{align}
    \gamma^t &=\sqrt{\frac{r_s}{r}}\,,\\
    \gamma^3 &= 1 - \frac{r_s}{r}\,.
\end{align}
We insert the expressions for $K$ and $k_{ab}$ into the equation for $\psi_Z$ and simplify. We obtain
\begin{align}
    \psi_Z &=\frac{r}{l(l+1)((l+2)(l-1)r + 3 r_s)}\Big[r^{-2} \qty((l+2)(l-1) r + 3 r_s) \bm x^h + 2 r \gamma^a \gamma^b h_{ab} - 2 \sqrt{\frac{r_s}{r}}  \partial_t \bm x^h\\
    &- 2 \qty(1 - \frac{r_s}{r}) \partial_r \bm x^h + \frac{4}{r}\qty(1 - \frac{r_s}{r}) \bm x^h + \frac{1}{\sqrt{l(l+1)}}\qty(4 r^2 \gamma^b \nabla_b (r^{-1} \gamma^a) j_a - 4 \Lambda \gamma^a j_a )\Big]\,.
\end{align}
In the appendix we showed that $\gamma$ satisfies the relation
\begin{align}
    \nabla_a \nabla_b \gamma = \frac{g_{ab}}{2r^2} r_s\,.
\end{align}
Inserting this into $\psi_Z$ and simplifying we get
\begin{align}
    \psi_Z = \frac{r}{l(l+1)((l+2)(l-1)r + 3 r_s)}\Big[&r^{-2}\qty((l^2 + l + 2) r - r_s) \bm x^h + 2 r \gamma^a \gamma^b h_{ab} - 2 \sqrt{\frac{r_s}{r}}  \partial_t \bm x^h\\
    &- 2 \qty(1 - \frac{r_s}{r}) \partial_r \bm x^h + \frac{4}{r}\qty(1 - \frac{r_s}{r}) \bm x^h - 2 \sqrt{l(l+1)}\gamma^a j_a\Big]\,.
\end{align}

Then, we still need to determine the time derivative of $\bm x^h$. 
As in the odd parity sector, we need to compute the Poisson bracket with the Hamiltonian. We have
\begin{align}
    \dot{\bm x}^h &= \qty{\bm x^h,\delta N \cdot {}^{(1)}Z^v + \delta N^3 \dot {}^{(1)}Z^h + r^{-2} j_t \cdot {}^{(1)}Z^e + N {}^{(2)}C_v + N^3{}^{(2)}C_h}\\
    &= -\frac{1}{2} \pi_\mu \delta N + 2 r \delta N^3 - \sqrt{l(l+1)} j_t + N^3 \partial_r \bm x^h - N \bm y^v - \frac{1}{4} \pi_\mu \bm x^v\,.
\end{align}
The $h_{ab}$ term is also expanded using the explicit form of $\gamma^a$. The calculation gives
\begin{equation}
    \gamma^a \gamma^b h_{ab} = \bm x^v + \frac{1}{2r} \pi_\mu \delta N^3 - \frac{1}{8 r^2} \pi_\mu^2 \delta N\,.
\end{equation}
Finally, combining all the results the Lagrangian master variable turns into
\begin{align}
    \psi_Z =&\frac{1}{l(l+1)((l+2)(l-1)r + 3 r_s)}\Big[\qty(\qty(l^2 + l + 2) - \frac{r_s}{r}) \bm x^h  - 2 r \partial_r \bm x^h+ \frac{1}{2}\pi_\mu \bm y^v + \frac{1}{8}\qty(\pi_\mu^2 + 16 r^2) \bm x^v\Big]\,.
\end{align}
Up to an $l$-dependent factor this is precisely the even parity variable $Q^e$ found in the Hamiltonian treatment. Thus, not only the wave equations match perfectly but the variables in the two approaches are exactly the same quantities. 

\section{Conclusion}

In this manuscript we revisited second order perturbations of spherically symmetric vacuum spacetimes
from the perspective of the reduced phase space  formalism and derived the reduced Hamiltonian. 
It depends on two pairs of physical degrees of freedom $(Q^o,P^o)$ and $(Q^e,P^e)$. 
In these variables the Hamiltonian decouples into the odd parity and even parity sectors at second order. 
The equations of motion derived from this Hamiltonian match precisely the ones found in a linearization of
Einstein's equations. Our result generalizes previous works in the Hamiltonian formalism by Moncrief 
\cite{6} to arbitrary spherically symmetric backgrounds and arrives at a much simplified version 
of the physical Hamiltonian found in \cite{7,8} upon performing suitable 
canonical transformations.

The analysis demonstrates the power of the reduced phase space formalism to solve the constraints order by order 
in a systematic way. In this paper we started with the first two orders. This gives two effective free scalar field 
theories with the Regge-Wheeler and Zerilli potential. 
In the future the results may be extended to higher order perturbation theory. Then, the theory will include 
interactions of the perturbations. 

Further extensions of the results are possible. So far only gravitational degrees of freedom were considered. 
It would be desirable to include matter fields from the Standard Model of particle physics. Of particular 
interest are electromagnetic fields and fermions. For the study of astrophysical black holes an application 
of the formalism to spacetimes with axial symmetry, such as the Kerr black hole is necessary. 

Additionally, a quantization of the physical Hamiltonian is of interest for the study of quantum gravity 
phenomena of black holes. We can use techniques derived for free quantum field theories on curved spacetimes
for the quadratic part of the Hamiltonian 
and apply usual perturbative quantum field theory techniques for its  interaction part.
The quantum theory provides tools to study the quantum stability of black holes. In particular this 
is important for a more careful study of the Hawking effect \cite{Hawking1975}. Especially the final 
stages of the evaporation process are not well understood. Quantum gravity is used to study this regime 
\cite{Ewing2020,Ewing2022}. Several proposals for the fate of the black hole such as the black to white hole 
transition \cite{Martin-Dussaud2019,Han2023} or remnants \cite{AHARONOV198751,Giddings1992} 
are discussed in the literature. The present formalism was developed to address these questions squarely
as it enables to go beyond the semiclassical Einstein equations and to explore the interior of 
black holes. See \cite{I} for an outline of our programme.

\section*{Acknowledgements}
JN thanks the Hanns-Seidel-Stiftung for financial and intellectual support.

\appendix

\section{Tensor Spherical Harmonics}
\label{sec:sphericalHarmonics}

In the main text of the manuscript we decompose tensor fields on the 2-sphere $S^2$ into tensor spherical harmonics. For that we briefly review their definition and collect some useful formulas. As before $S^2$ is equipped with the usual metric given by $\Omega_{AB}$ and the torsion-free, metric-compatible covariant derivative is denoted by $D_A$. Furthermore, we use the two-dimensional Levi-Civita-pseudotensor $\epsilon_{AB}$.

The scalar spherical harmonics $Y_{lm}$ are complex valued functions on $S^2$. For the computations we choose a real version denoted by $L_{lm}$. They are eigenfunctions of the Laplace operator on the sphere, i.e. they satisfy $\Omega^{AB}D_A D_B L_{lm} = - l(l+1) L_{lm}$. The functions satisfy the orthogonality relation
\begin{equation}
    \int_{S^2} \sqrt{\Omega}\, L_{l_1 m_1} L_{l_2 m_2} = \delta_{l_1 l_2} \delta_{m_1 m_2}\,.
\end{equation}
The tensor harmonics are then defined in terms of these functions. We have
\begin{align}
    L^A_{e,lm} &:= \frac{1}{\sqrt{l(l+1)}} D^A L_{lm}\\
    L^A_{o,lm} &:= \frac{1}{\sqrt{l(l+1)}} \varepsilon^{AB} D_B L_{lm}\\
    L^{AB}_{e,lm} &:= \sqrt{\frac{2}{(l-1)l(l+1)(l+2)}} \qty(D^A D^B + \frac{1}{2}l(l+1)\Omega^{AB}) L_{lm}\\
    L^{AB}_{o,lm} &:= \sqrt{\frac{2}{(l-1)(l+2)}} D^{(A}L^{B)}_{o,lm}\,.
\end{align}
Here the correct normalizations were chosen so that we have
\begin{align}
    &\int_{S^2} \sqrt{\Omega}\, \Omega_{AB} L^A_{I_1,l_1 m_1} L^B_{I_2,l_2 m_2} = \delta_{I_1 I_2}\delta_{l_1 l_2} \delta_{m_1 m_2}\,,\\
    &\int_{S^2} \sqrt{\Omega}\, \Omega_{AC}\Omega_{BD} L^{AB}_{I_1,l_1 m_1} L^{CD}_{I_2,l_2 m_2} = \delta_{I_1 I_2}\delta_{l_1 l_2} \delta_{m_1 m_2}\,,
\end{align}
where $I_1,I_2$ are labeling the parity of the harmonics ($e$ or $o$). 

We also present the following identities which are very useful in explicit computations:
\begin{align}
    D_M D_B L^M_{o,lm} &= L_{B,o,lm}\\
    D^M D_M L^A_{o,lm} &= (1 - l (l+1)) L^A_{o,lm}\\
    D_M L^{MA}_{o,lm} &= - \sqrt{\frac{(l+2)(l-1)}{2}} L^A_{o,lm}\\
    D_M D^{(A} L^{B)M}_{o,lm} &= \frac{1}{2}\qty(6 - l(l+1)) L^{AB}_{o,lm}\\
    D^M D_M L^{AB}_{o,lm} &= (4 - l(l+1)) L^{AB}_{o,lm}\\
    D_M D_B L^M_{e,lm} &= (1 - l (l+1))L_{B,e,lm}\\
    D^M D_M L^A_{e,lm} &= (1 - l(l+1))L^A_{e,lm}\\
    D_M L^{MA}_{e,lm} &= - \sqrt{\frac{(l+2)(l-1)}{2}} L^A_{e,lm}\\
    D^M D_M L^{AB}_{e,lm} &= \qty(4 - l(l+1)) L^{AB}_{e,lm}\\
    D_M D^A L^{MB}_{e,lm} &= \frac{1}{2}\qty(6 - l(l+1)) L^{AB}_{e,lm} + \frac{1}{2} \sqrt{\frac{(l-1)l(l+1)(l+2)}{2}} \Omega^{AB} L_{lm}
\end{align}

\section{Expansion of the Diffeomorphism and Hamiltonian Constraints to Second order}
\label{sec:Expansion2ndOrder}

In this appendix we present the expansion of the diffeomorphism and Hamiltonian constraint to second order. As the spherically symmetric background spacetime we use the Gullstrand-Painlevé coordinates explained in the main text. In these coordinates the spatial metric is flat, i.e. $m_{33}=1$ and $m_{AB} = r^2 \Omega_{AB}$. The Christoffel symbols take the form 
\begin{align}
    \Gamma^3_{AB} &= - r \Omega_{AB}\\
    \Gamma^A_{rB} &= r^{-1} \delta^A_B\,.
\end{align}
$\Gamma^A_{BC}$ is the Christoffel symbols on the 2-sphere. All other components of the Christoffel symbol vanish. The Ricci tensor, Ricci scalar and Einstein tensor of the induced metric $m_{\mu \nu}$ vanish. 

In the following we split the metric $m_{\mu \nu}$ into background and perturbations. We use the notation $m_{\mu \nu} = \overline m_{\mu \nu} + \delta m_{\mu \nu}$ and $W^{\mu \nu} = \overline W^{\mu \nu} + \delta W^{\mu \nu}$, where the barred quantities are the background and the 
variables with $\delta$ the perturbations. In particular $\Box$ denotes the 
background Laplacian.
In the computations we raise and lower indices with the background 
metric $\overline m_{\mu \nu}$.  
For later convenience we recall the perturbation 
formulae for various quantities to second order. We have

\begin{align}
    \begin{split}
    m^{\mu \nu} =& \overline m^{\mu \nu} - \delta m^{\mu \nu} + \delta m^{\mu \rho} \delta m^{\nu \sigma} \overline m_{\rho \sigma}\\
    \sqrt{m} =& \sqrt{\overline m}\qty(1 + \frac{1}{2}\delta m^\mu{}_\mu + \frac{1}{8} \qty((\delta m^\mu{}_\mu)^2 - 2 \delta m^{\mu \nu}\delta m_{\mu \nu}))\\
    \sqrt{m}^{-1} =& \sqrt{\overline m}^{-1}\qty(1 - \frac{1}{2} \delta m^\mu{}_\mu +\frac{1}{8} \qty((\delta m^\mu{}_\mu)^2 + 2 \delta m^{\mu \nu}\delta m_{\mu \nu}))\\
    \Gamma^\mu_{\nu \rho} =& \overline \Gamma^\mu_{\nu \rho} + \frac{1}{2} \overline m^{\mu \sigma}\qty(\nabla_\nu \delta m_{\rho \sigma} + \nabla_\rho \delta m_{\nu \sigma} - \nabla_\sigma \delta m_{\nu \rho}) - \frac{1}{2} \delta m^{\mu \sigma} \qty(\nabla_\nu \delta m_{\rho \sigma} + \nabla_\rho \delta m_{\nu \sigma} - \nabla_\sigma \delta m_{\nu \rho})\\
    R_{\mu \nu} =& \frac{1}{2}\left(2 \overline m^{\rho \sigma} \nabla_\rho \nabla_{(\mu}\delta m_{\nu)\sigma} - \square \delta m_{\mu \nu} -  \nabla_\mu \nabla_\nu \delta m^\rho{}_\rho\right) - \frac{1}{2}\nabla_\rho \qty( \delta m^{\rho \sigma}\qty(2 \nabla_{(\mu} \delta m_{\nu) \sigma} - \nabla_{\sigma} \delta m_{\mu \nu}))\\
    &+ \frac{1}{2}\nabla_\nu \qty(\delta m^{\rho \sigma} \nabla_\mu \delta m_{\rho \sigma}) + \frac{1}{4} \nabla^\sigma \delta m^\rho{}_\rho \left( 2 \nabla_{(\mu} \delta m_{\nu) \sigma} - \nabla_\sigma \delta m_{\mu \nu}\right) \\
    &- \frac{1}{4}\left(2 \nabla_{(\rho}\delta m_{\nu)\sigma} - \nabla_\sigma \delta m_{\mu \rho}\right)\left(2 \nabla_{(\nu}\delta m_{\alpha) \beta} - \nabla_\beta \delta m_{\alpha \nu}\right)\overline m^{\alpha \sigma} \overline m^{\rho \beta}\\
    R =& \nabla_\mu \nabla_\nu \delta m^{\mu \nu} - \square \delta m^\mu{}_\mu\\
    &+ \delta m^{\mu \nu}\left(\square \delta m_{\mu \nu} +  \nabla_\mu \nabla_\nu \delta m^\rho{}_\rho- 2 \overline m^{\rho \sigma} \nabla_{(\rho} \nabla_{\mu)}\delta m_{\nu \sigma}\right) - \nabla_\rho \delta m^{\rho \sigma} \nabla^{\mu} \delta m_{\mu \sigma}\\
    &+\nabla_\rho \delta m^{\rho \sigma}\nabla_{\sigma} \delta m^\mu{}_{\mu} + \frac{3}{4}\nabla^\mu \delta m^{\rho \sigma} \nabla_\mu \delta m_{\rho \sigma} - \frac{1}{4} \nabla^\sigma \delta m^\rho{}_\rho \nabla_\sigma \delta m^{\mu}{}_{\mu}- \frac{1}{2} \nabla^{\rho}\delta m_{\mu \sigma} \nabla^\sigma\delta m^\mu{}_{\rho}
    \end{split}
\end{align}

We use these formulas and expand the diffeomorphism constraint $V_\mu$ and the Hamiltonian constraint $V_0$ to second order in $\delta m_{\mu \nu}$ and  $\delta W^{\mu \nu}$. Then, we split the spatial indices into the part containing indices of the sphere $A,B,\dots$ and the part containing the index $3$. 

The first order correction to the diffeomorphism constraint is given by

\begin{equation}
    {}^{(1)}V_\mu = - 2 \nabla_\rho \qty(\delta m_{\mu\nu} \overline W^{\nu \rho} + \overline m_{\mu \nu} \delta W^{\nu \rho}) + \overline W^{\nu \rho} \nabla_\mu \delta m_{\nu \rho}
\end{equation}
Performing the explicit splitting into the radial and spherical parts we obtain

\begin{align}
    \begin{split}
    {}^{(1)}V_3 =& -2 \partial_r \qty(\delta W^{33})- 2 D_A \delta W^{3A} + 2 r \Omega_{AB}\delta W^{AB} - \sqrt{\Omega} \pi_\mu' \delta m_{33}\\
    &- \frac{\sqrt{\Omega}\pi_\mu}{2} \partial_r \delta m_{33} - \sqrt{\Omega} \frac{\pi_\lambda}{2 r^2} D^A \delta m_{3A} + \sqrt{\Omega}\frac{\pi_\lambda}{4 r^2} \Omega^{AB} \partial_r \delta m_{AB}
    \end{split}\\
    \begin{split}
    {}^{(1)}V_A =& -2 \Omega_{AB} \partial_r \qty(r^2 \delta W^{3 B}) - 2 r^2 \Omega_{AB} D_C \delta W^{BC}+ \frac{\sqrt{\Omega} \pi_\mu}{2} D_A \delta m_{33}\\
    &- \partial_r \qty(\delta m_{3A} \sqrt{\Omega}\pi_\mu)+ \frac{\sqrt{\Omega} \pi_\lambda}{4 r^2} \qty(\Omega^{CD}  D_A \delta m_{CD} - 2 D^B \delta m_{AB})
    \end{split}
\end{align}
Here and in the following we assume that indices $A,B,\dots$ will be moved with $\Omega_{AB}$. 

The second order correction to the diffeomorphism constraint is
\begin{equation}
    {}^{(2)}V_\mu = - 2 \nabla_\rho \qty(\delta m_{\mu \nu} \delta W^{\nu \rho}) + \delta W^{\nu \rho} \nabla_\mu \delta m_{\nu \rho} 
\end{equation}
For the computation in the main text we only need the radial component of this constraint. It is
\begin{align}
    \begin{split}
    {}^{(2)}V_3 =& \delta W^{33} \partial_r \delta m_{33} - 2 \partial_r \qty(\delta W^{33} \delta m_{33}) - 2 \delta m_{3A} \partial_r \delta W^{3A}\\
    &- 2 D_A \qty(\delta m_{33} \delta W^{3A}) + \delta W^{AB} \partial_r \delta m_{AB} - 2 D_A \qty(\delta m_{3B} \delta W^{AB})
    \end{split}
\end{align}

This completes the diffeomorphism constraint. We now move to the Hamiltonian constraint $V_0$. The first order correction is given by
\begin{align}
    \begin{split}
    {}^{(1)}V_0 =&\frac{1}{\sqrt{\overline m}}\qty(\overline W^{\rho \sigma} \overline m_{\mu \rho} \overline m_{\nu \sigma} - \frac{1}{2} \overline m_{\mu \nu} \overline W^{\rho \sigma}\overline m_{\rho \sigma})\qty(2 \delta W^{\mu \nu}+ 2 \overline m^{\mu \alpha}  \delta m_{\alpha \beta} \overline W^{\beta \nu}- \frac{1}{2} \overline m^{\alpha \beta} \delta m_{\alpha \beta} \overline W^{\mu \nu})\\
    & + \sqrt{\overline m}\qty(- \nabla_\mu \nabla_\nu \delta m_{\rho \sigma} \overline m^{\mu \rho} \overline m^{\nu \sigma} + \square \delta m_{\mu \nu} \overline m^{\mu \nu})
    \end{split}
\end{align}
We split all the indices into their radial and angular parts and get
\begin{align}
    {}^{(1)}V_0 =& \frac{1}{2r^2}\qty(\pi_\mu - \pi_\lambda)  \delta W^{33} - \frac{1}{2} \pi_\mu  \delta W^{AB} \Omega_{AB} - \frac{1}{16 r^4} \sqrt{\Omega} \pi_\mu^2 \Omega^{AB} \delta m_{AB} + \frac{1}{16 r^2} \sqrt{\Omega} (3 \pi_\mu^2 - 2 \pi_\mu \pi_\lambda)  \delta m_{33}\nonumber\\
    &+ \sqrt{\Omega} \Big[(D_A D^A \delta m_{33}  - 2 r \partial_r \delta m_{33} -  2 \delta m_{33}) -2 (\partial_r + r^{-1}) D^A \delta m_{3A}\\
    &+ r^{-2} \qty(D_A D^A \delta m_{CD} \Omega^{CD} - D^A D^B \delta m_{AB}) + \qty( \partial_r^2  - r^{-1} \partial_r + r^{-2}) \delta m_{AB}\Omega^{AB}\Big]\nonumber
\end{align}

The second order corrections to the Hamiltonian constraint are given by
\begin{align}
    \label{eq:Ham2ndOrder}
    {}^{(2)}V_0 =&\frac{1}{r^2 \sqrt{\Omega}}\Big[\delta W^{\mu \nu} \delta W^{\rho \sigma} \qty(\overline m_{\mu \rho} \overline m_{\nu \sigma} - \frac{1}{2} \overline m_{\mu \nu}\overline m_{\rho \sigma}) + \delta W^{\mu \nu} \overline W^{\rho \sigma} \qty(4 \delta m_{\mu \rho} \overline m_{\nu \sigma} - \qty(\delta m_{\mu \nu} \overline m_{\rho \sigma} + \delta m_{\rho \sigma} \overline m_{\mu \nu}))\nonumber\\
    &+\overline W^{\mu \nu} \overline W^{\rho \sigma}\qty(\delta m_{\mu \rho} \delta m_{\nu \sigma} - \frac{1}{2} \delta m_{\mu \nu} \delta m_{\rho \sigma}) + \frac{1}{8}\qty(2 \delta m^{\mu \nu} \delta m_{\mu \nu} + (\delta m^{\mu}{}_{\mu})^2 )\qty(\overline W^{\rho \sigma}\overline W_{\rho \sigma} - \frac{1}{2} (\overline W^\rho{}_\rho)^2)\nonumber\\
    &+ \frac{1}{2} \delta m^\rho_\rho\qty( \delta W^\mu{}_\mu W^\nu{}_\nu + W^{\sigma}{}_\sigma W^{\mu \nu}  \delta m_{\mu \nu} - 2 W^{\mu \nu}W^{\sigma \tau} \delta m_{\mu \sigma} m_{\nu \tau} -2 \delta W^{\mu \nu} W_{\mu \nu} )\Big]\\
    &- r^2 \sqrt{\Omega}\Big[\nabla_\mu \delta m_{\nu \rho} \nabla_\sigma \delta m_{\alpha \beta} \overline m_{(1)}^{\mu \nu \rho \sigma \alpha \beta} + \delta m_{\mu \nu} \nabla_\rho \nabla_\sigma \delta m_{\alpha \beta} \overline m_{(2)}^{\mu \nu \rho \sigma \alpha \beta}\Big]\nonumber
\end{align}
where we raised and lowered indices with the metric $\overline m_{\mu \nu}$ and defined the following tensors:
\begin{align}
    \overline m_{(1)}^{\mu \nu \rho \sigma \alpha \beta}:=&-\overline m^{\mu \nu} \overline m^{\rho \beta} \overline m^{\sigma \alpha} + \overline m^{\mu \nu} \overline m^{\rho \sigma} \overline m^{\alpha \beta} + \frac{3}{4} \overline m^{\mu \sigma} \overline m^{\nu \alpha} \overline m^{\rho \beta} - \frac{1}{4} \overline m^{\mu \sigma} \overline m^{\nu \rho} \overline m^{\alpha \beta} - \frac{1}{2} \overline m^{\mu \beta} \overline m^{\nu \alpha} \overline m^{\rho \sigma}\\
    \begin{split}
    \overline m_{(2)}^{\mu \nu \rho \sigma \alpha \beta} :=& \frac{1}{2} \overline m^{\mu \nu} \overline m^{\rho \alpha} \overline m^{\sigma \beta} - \frac{1}{2} \overline m^{\mu \nu} \overline m^{\rho \sigma} \overline m^{\alpha \beta} +  \overline m^{\mu \alpha}\overline m^{\nu \beta} \overline m^{\rho \sigma} + \overline m^{\mu \rho} \overline m^{\nu \sigma} \overline m^{\alpha \beta} - \overline m^{\mu \sigma} \overline m^{\nu \alpha} \overline m^{\rho \beta}\\
    &- \overline m^{\mu \rho} \overline m^{\nu \alpha} \overline m^{\sigma \beta}\,.
    \end{split}
\end{align}

For the computation it is convenient to expand the second covariant derivatives into the radial and angular components explicitly. We have
\begin{align}
    \begin{split}
    \nabla_r \nabla_r \delta m_{33} &= \partial_r^2 \delta m_{33}\\
    \nabla_r \nabla_r \delta m_{3A} &= (\partial_r - r^{-1})(\partial_r - r^{-1}) \delta m_{3A}\\
    \nabla_r \nabla_r \delta m_{AB} &= (\partial_r - 2 r^{-1})(\partial_r - 2 r^{-1}) \delta m_{AB}\\
    \nabla_r \nabla_A \delta m_{33} &=(\partial_r - r^{-1})\qty(D_A \delta m_{33} - 2 r^{-1} \delta m_{3A})\\
    \nabla_r \nabla_A \delta m_{3B} &=(\partial_r - 2 r^{-1})\qty( D_A \delta m_{3B} - r^{-1} \delta m_{AB} + r \Omega_{AB} \delta m_{33})\\
    \nabla_r \nabla_A \delta m_{BC} &=(\partial_r - 3 r^{-1})\qty(D_A \delta m_{BC} + r \qty(\Omega_{AB} \delta m_{3C} + \Omega_{AC} \delta m_{3B}))\\
    \nabla_A \nabla_r \delta m_{33} &=D_A (\partial_r - r^{-1}) \delta m_{33} - 2 r^{-1} (\partial_r - 2 r^{-1}) \delta m_{3A}\\
    \nabla_A \nabla_r \delta m_{rB} &=r \Omega_{AB} (\partial_r - r^{-1}) \delta m_{33} + D_A (\partial_r -2 r^{-1}) \delta m_{3B} - r^{-1} (\partial_r - 3 r^{-1}) \delta m_{AB}\\
    \nabla_A \nabla_r \delta m_{BC} &=D_A (\partial_r - 3 r^{-1} \delta m_{BC} + r \qty(\Omega_{AB} (\partial_r - 2 r^{-1}) \delta m_{3C} + \Omega_{AC} (\partial_r - 2 r^{-1}) \delta m_{3B})\\
    \nabla_A \nabla_B \delta m_{33} &=D_A D_B \delta m_{33} + r \Omega_{AB} (\partial_r - 2 r^{-1}) \delta m_{33} - 4 r^{-1} D_{(A}\delta m_{B)3} + 2 r^{-2} \delta m_{AB}\\
    \nabla_A \nabla_B \delta m_{3C} &= 2 r \Omega_{C(A}D_{B)} \delta m_{33} + D_A D_B \delta m_{3C} + r \Omega_{AB} (\partial_r - 2 r^{-2}) \delta m_{3C}\\
    &- 2 \Omega_{AC} \delta m_{3B} - \Omega_{BC} \delta m_{3A} - 2 r^{-1} D_{(A} \delta m_{B)C}\\
    \nabla_A \nabla_B \delta m_{CD} &=r^2 \qty(\Omega_{BC} \Omega_{AD}+ \Omega_{BD} \Omega_{AC}) \delta m_{33}+ 2 r \qty(\Omega_{C(A}D_{B)}\delta m_{rD} + \Omega_{D(A} D_{B)} \delta m_{3C})\\
    &+ D_A D_B \delta m_{CD}+ r \qty(\Omega_{AB}  (\partial_r - 2 r^{-1}) \delta m_{CD} - 2 r^{-1} \Omega_{A(C}\delta m_{D)B})
    \end{split}
\end{align}

We now give the decomposition of the second order Hamiltonian constraint into the radial and angular components in three steps. First we consider the momentum contributions which are the first three lines \eqref{eq:Ham2ndOrder}. They are
\begin{align}
    \begin{split}
    &\frac{1}{2} (\delta W^{33})^2 + r^2 \Omega_{AB} \qty(2 \delta W^{3A} \delta P^{3B} - \delta W^{33} \delta W^{AB}) + r^4 \qty(\delta W^{AB} \delta W_{AB} - \frac{1}{2} (\delta W^{AB}\Omega_{AB})^2) \\
    &+ \frac{1}{4}(3 \pi_\mu - \pi_\lambda) \delta W^{33} \delta m_{33}  - \frac{\pi_\mu}{4 r^2} \delta W^{33} \Omega^{AB} \delta m_{AB}  + \pi_\mu \delta W^{3 A} \delta m_{3A} - \frac{\pi_\mu}{4} r^2  \Omega_{AB} \delta W^{AB} \delta m_{33} \\
    &- \frac{1}{2}\delta W^{AB} \delta m_{AB} (\pi_\mu - \pi_\lambda)+ \frac{1}{4}(\pi_\mu - \pi_\lambda) \Omega_{AB}\Omega^{CD} \delta W^{AB} \delta m_{CD}\\
    &+\frac{1}{64}\qty( 3 \pi_\mu^2 + 2 \pi_\mu \pi_\lambda)(\delta m_{33})^2 + \frac{1} {16 r^2}\qty(\pi_\mu^2 + 2 \pi_\mu \pi_\lambda) \Omega^{AB}\delta m_{rA}\delta m_{rB} - \frac{3}{32 r^2} \pi_\mu^2 \delta m_{33} \Omega^{AB} \delta m_{AB}\\
    &+  \frac{1}{32 r^4}\qty(\pi_\mu^2 - 2\pi_\mu \pi_\lambda + 2 \pi_\lambda^2) \Omega^{AB} \Omega^{CD} \delta m_{AC} \delta m_{BD} +\frac{1}{64 r^4}\qty(\pi_\mu^2  + 2 \pi_\mu \pi_\lambda - 2 \pi_\lambda^2) \Omega^{AB} \Omega^{CD} \delta _{AB} \delta m_{CD}
    \end{split}
\end{align}
The next term in \eqref{eq:Ham2ndOrder} involves first order derivatives. The splitting of this contribution is
\begin{equation}
    \resizebox{0.9\textwidth}{!}{$\begin{aligned}
    &- r^2 \sqrt{\Omega}\nabla_\mu \delta m_{\nu \rho} \nabla_\sigma \delta m_{\alpha \beta} \overline m_{(1)}^{\mu \nu \rho \sigma \alpha \beta}=-\sqrt{\Omega} \Big[ \frac{1}{2}\partial_r \delta m_{33} \partial_r \delta m_{AB}\Omega^{AB} - \partial_r \delta m_{33}\qty(D^A \delta m_{3A} + 2 r \delta m_{33} ) \\
    &+\frac{1}{2} \qty(D_A \delta m_{33} - 2 r^{-1} \delta m_{3 A})\qty(D_B \delta m_{33} - 2 r^{-1} \delta m_{3 B})\Omega^{AB}\\
    &+\frac{1}{r^2} \qty(D_A \delta m_{BC} + r \qty(\Omega_{AB} \delta m_{3C} + \Omega_{AC} \delta m_{3B}))\qty(D_D \delta m_{33} - 2 r^{-1} \delta m_{3D}) \qty(\Omega^{AB} \Omega^{CD} - \frac{1}{2} \Omega^{AD} \Omega^{BC})\\
    &+\frac{1}{r^2}\qty(D_A \delta m_{BC} + r \qty(\Omega_{AB} \delta m_{3C} + \Omega_{AC} \delta m_{3B}))(\partial_r - r^{-1})\delta m_{3D}(\Omega^{AD}\Omega^{BC}-2\Omega^{AB}\Omega^{CD})\\
    &+\frac{1}{r^2}\qty(D_A \delta m_{3B} - r^{-1} \delta m_{AB} + r \Omega_{AB} \delta m_{33})\qty(D_C \delta m_{3D} - r^{-1} \delta m_{CD} + r \Omega_{CD} \delta m_{33})\times\\
    &~~~~~~~~~~~~~~~~~~~~~~~~~~~~~~~~~~~~~~~~\times\qty(\frac{3}{2} \Omega^{AC} \Omega^{BD} - \frac{1}{2} \Omega^{AD}\Omega^{BC} - \Omega^{AB} \Omega^{CD})\\
    &+\frac{1}{r^2}(\partial_r - 2 r^{-1})\delta m_{AB}\qty(D_C \delta m_{3D} - r^{-1} \delta m_{CD} + r \Omega_{CD} \delta m_{33})(\Omega^{AB} \Omega^{CD} - \Omega^{AD} \Omega^{BC})\\
    &+\frac{1}{4r^2}(\partial_r -2 r^{-1})\delta m_{AB} (\partial_r - 2 r^{-1})\delta m_{CD}(3 \Omega^{AC} \Omega^{BD}-\Omega^{AB}\Omega^{CD})\\
    &+ \frac{1}{r^4} \qty(D_A \delta m_{BC} + r(\Omega_{AB} \delta m_{rC} + \Omega_{AC}\delta m_{3B} ))\qty(D_D \delta m_{EF} + r (\Omega_{DE} \delta m_{3F} + \Omega_{DF}\delta m_{3E} ))\Omega_{(1)}^{ABCDEF}\Big]
    \end{aligned}$}
\end{equation}
Finally we get the last term in \eqref{eq:Ham2ndOrder}. It consists of the contributions with second derivatives. 
\begin{equation}
    \resizebox{0.9\textwidth}{!}{$\begin{aligned}
    &- r^2 \sqrt{\Omega} \delta m_{\mu \nu} \nabla_\rho \nabla_\sigma \delta m_{\alpha \beta} \overline m_{(2)}^{\mu \nu \rho \sigma \alpha \beta}\\
    &=-\sqrt{\Omega}\delta m_{33} \Big[\frac{1}{2}\nabla_r \nabla_r \delta m_{AB} \Omega^{AB} - \frac{1}{2} \nabla_r \nabla_A \delta m_{3B} \Omega^{AB}  - \frac{1}{2} \nabla_A \nabla_r \delta m_{3B} \Omega^{AB} +\frac{1}{2} \nabla_A \nabla_B \delta m_{33} \Omega^{AB} \Big]\\
    &- \sqrt{\Omega} r^{-2} \Big[ \frac{1}{2}\delta m_{33} \nabla_A \nabla_B \delta m_{CD}(\Omega^{AC}\Omega^{BD} - \Omega^{AB}\Omega^{CD}) + \delta m_{3A} \nabla_r \nabla_B \delta m_{CD}(\Omega^{AB} \Omega^{CD}-\Omega^{AC}\Omega^{BD})\\
    &+\delta m_{3A} \nabla_B \nabla_C \delta m_{3D}(2\Omega^{AD}\Omega^{BC} - \Omega^{AC} \Omega^{BD} - \Omega^{AB} \Omega^{CD}) + \delta m_{3A} \nabla_B \nabla_r \delta m_{CD}(\Omega^{AB} \Omega^{CD} - \Omega^{AC} \Omega^{BD})\\
    &+\delta m_{AB} \nabla_r \nabla_r \delta m_{CD}\qty(\Omega^{AC} \Omega^{BD} - \frac{1}{2}\Omega^{AB}\Omega^{CD}) + \delta m_{AB} \nabla_r \nabla_C \delta m_{rD}\qty(\frac{1}{2} \Omega^{AB} \Omega^{CD} - \Omega^{AC} \Omega^{BD})\\
    &+ \delta m_{AB} \nabla_C \nabla_r \delta m_{3D}\qty(\frac{1}{2} \Omega^{AB} \Omega^{CD} - \Omega^{AC} \Omega^{BD}) + \delta m_{AB} \nabla_C \nabla_D \delta m_{rr}\qty(\Omega^{AC} \Omega^{BD} - \frac{1}{2}\Omega^{AB} \Omega^{CD})\Big]\\
    &-\sqrt{\Omega} r^{-4} \delta m_{AB} \nabla_C \nabla_D \delta m_{EF} \Omega_{(2)}^{ABCDEF}
    \end{aligned}$}
\end{equation}
We now have to carefully insert the definition of the perturbed metric and perturbed momentum. After integrating out the spherical coordinates we obtain the result shown in the main text.

\section{Boundary term even parity}
\label{sec:EvenParityBoundaryTerm}
The boundary term for the even parity perturbations is given by
\begin{align}
    &\frac{r+3 r_s}{2 r^2} (q_2')^2 + \frac{r_s -\left(l^2+l+2\right) r}{r^3} q_2' q_2 + \frac{2(r+r_s)}{r}q_2' q_1 - 3 \frac{\sqrt{r r_s}}{r^2}q_2' p_1 -\frac{\left(3 r-r_s\right) \left(\left(l^2+l+2\right) r-r_s\right)}{2 r^2 \left(r_s+r\right)} q_2 q_1 \nonumber\\
    &+ \frac{\sqrt{r r_s} \left(r_s+3 r\right) \left(\left(l^2+l+2\right) r-r_s\right)}{r^3 \left(r_s+r\right){}^2}q_2 p_1 +  \frac{-r^2+6 r r_s +3 r_s^2}{2 r (r+r_s)^2} p_1^2 -\frac{2 r_s (3 r+r_s)}{\sqrt{r r_s} (r+r_s)} p_1 q_1 \nonumber\\
    &+\frac{1}{2 l (l+1) \left(2 \left(l^2+l-2\right) r+3 r_s\right){}^2}\Big[3 \left(19 l^2+19 l+10\right) r r_s^2+\left(29 l^4+58 l^3-19 l^2-48 l-20\right) r^2 r_s\nonumber\\
    &~~~~~~~~~~~~~~~~~~~~~+\left(3 l^6+9 l^5-l^4-17 l^3+2 l^2+12 l-8\right) r^3+27r_s^3\Big] (q_1)^2\nonumber\\
   &-\frac{1}{8 r^4 \left(r_s+r\right){}^2}\Big[\left(3 l^2+3 l+2\right) r_s^3+l \left(l^3+2 l^2+13 l+12\right) r r_s^2\\
   &~~~~~~~~~~~~~~~~~~~~~-\left(\left(3 l^4+6 l^3+13 l^2+10 l+16\right) r^3\right)+\left(2 l^4+4
   l^3+25 l^2+23 l+18\right) r^2 r_s\Big] (q_2)^2\nonumber\\
   &-\frac{(l+2)(l-1) r^2+6 r r_s + 3 r_s^2}{2 r (r+r_s) \left((l+2)(l-1) r+3 r_s \right)} (Q^e)^2\nonumber\\
   &+\frac{r_s}{8 \left(l^2+l-2\right) r^2 (r+r_s)^2 \left(\left(l^2+l-2\right) r+3 r_s\right)}\Big[6 \left(5 l^2+5 l-31\right) r_s^3\nonumber\\
   &~~~~~~~~~~~~~~~~~~~~~~+\left(l^2+l-2\right)^2
   \left(l^4+2 l^3+13 l^2+12 l+3\right) r^3+3 \left(9 l^4+18 l^3+5 l^2-4 l+35\right) r r_s^2\nonumber\\
   &~~~~~~~~~~~~~~~~~~~~~~+ 2 \left(5 l^6+15 l^5+28 l^4+31 l^3-15 l^2-28 l-36\right) r^2 r_s \Big] (Q^e)^2\nonumber
\end{align}

\section{Boundary term for proof of weak gauge invariance}
\label{sec:BoundaryTermProofGaugeInv}
We present the boundary term of $\pi_\mu^{(2)}$ split into different contributions. We will call these contributions $A_1,\dots A_4$. The first one, $A_1$. involves the terms containing $q_2$. Parts of it are already present in the boundary term in section \ref{sec:EvenParityBoundaryTerm}. We have
\begin{align*}
        A_1=&- \frac{r + 3 r_s}{2 r^2} (q_2')^2 + \frac{r_s - (l^2+l+2)r}{r^3}q_2 q_2' - \frac{2(r+r_s)}{r} q_1 q_2' + \frac{3 \sqrt{r rs}}{r^2} p_1 q_2' + \frac{\sqrt{l(l+1)}}{r} \bm x^e q_2' - \sqrt{\frac{r_s}{r}} p_2 q_2 \\
        &+ \frac{(3r - r_s)((2 + l + l^2)r - r_s)}{2 r^2(r+r_s)}q_1 q_2 - \frac{\sqrt{r r_s}((2+l+l^2)r - r_s)(3r + r_s)}{r^3(r+r_s)^2} p_1 q_2\\
        &+\frac{\sqrt{l(l+1)}(r_s - (l^2+l+2) r)}{2 r^2 (r+r_s)} \bm x^e q_2 + \frac{\sqrt{2(l-1)l(l+1)(l+2)}}{4 r^3} \bm X^e q_2\tag{\stepcounter{equation}\theequation}\\
        &+\frac{1}{8 r^4 (r+r_s)^2}\Big[l(l+1) \left(l^2+l+12\right)r r_s^2 -\left((l (l+1) (2 l (l+1)+23)+18)r^2 r_s\right)\\
        &~~~~~~~~~~~~~~~~~~~~~~-(l (l+1) (3 l (l+1)+10)+16) r^3 + (3 l (l+1)+2)  r_s^3\Big]q_2^2\\
\end{align*}
Next, $A_2$ involves the everything with the variable $\bm X^e$ that is not part of $A_1$:
\begin{align*}
    A_2=&- \frac{1}{r^2}\bm X^e \bm X^e{}' + \frac{r_s \left(\left(l^2+l+6\right) r+3 r_s\right)}{\sqrt{2} \sqrt{l^2+l-2} r^2 (r+ r_s)} \bm x^e \bm X^e{}'\\
    &+\frac{\sqrt{r r_s}}{\sqrt{2(l+2)(l+1)l(l-1)} l(l+1) r (r+r_s)^2 \Lambda} \Big[(2 l \left(l^3+2 l^2-27 l-28\right) r^2 r_s\\
    &~~~~~~~~~~~~~+\left(7 l^4+14 l^3+5 l^2-2 l-3\right) r^3-\left(l^4+2 l^3+23 l^2+22 l-117\right) r r_s^2+54 r_s^3\Big]\bm X^e p_2'\\
   &+\frac{\sqrt{r r_s}}{2 r^3 \sqrt{2(l-1) l (l+1) (l+2)} (r+r_s)^3} \Big[2 l (l+1) \left(l^2+l-28\right) r^2 r_s\\
   &~~~~~~~~~~~~~- \left(l (l+1)
   \left(l^2+l+22\right)-117\right) r r_s^2+(l (l+1) (7 l (l+1)-2)-3) r^3+54 r_s^3\Big] O_{p_1} \bm X^e\\
   &+\frac{\sqrt{r r_s}}{4 l(l+1)\sqrt{2(l+2)(l+1)l(l-1))}r^3
   \left(r_s+r\right){}^3  \Lambda^2} \Big[-27 \left(l^2+l-54\right) r_s^5\\
   &~~~~~~~~~~~~~+\left(l (l+1) \left(l (l+1) \left(l^2+l-5\right) (3 l (l+1)+52)+579\right)-1017\right) r^3
   r_s^2\\
   &~~~~~~~~~~~~~-(l-1) (l+2) (l (l+1) (l (l+1) (5 l (l+1)-29)+11)+12)
   r^5\\
   &~~~~~~~~~~~~~+(l (l+1) (l (l+1) (l (l+1) (2 l (l+1)+95)-334)+404)-12) r^4 r_s\tag{\stepcounter{equation}\theequation}\\
   &~~~~~~~~~~~~~+(l (l+1) (l (l+1) (5 l
   (l+1)-86)-757)+36) r^2 r_s^3-9 (l (l+1) (3 l (l+1)+41)-375) r r_s^4\Big]p_2 \bm X^e\\
   &-\frac{r_s\left(\left(l^2+l+6\right) r+3 r_s\right)}{2 l (l+1) \sqrt{2(l+2)(l-1)} r^3 \left(r_s+r\right){}^3 \Lambda} \Big[2 l (l+1) \left(l^2+l-28\right)r^2 r_s\\
   &~~~~~~~~~~~~~+\left(-l (l+1) \left(l^2+l+22\right) r r_s^2+54 r_s^3+117 r r_s^2\right)+(l (l+1) (7 l (l+1)-2)-3)
   r^3\Big]\bm x^e{}' \bm X^e\\
   &+\frac{1}{16 (l-1) l (l+1) (l+2) r^4 \left(r_s+r\right){}^4}\Big[(l-1) l (l+1) (l+2) (l (l+1) (3 l (l+1)-8)+24) r^5\\
   &~~~~~~~~~~~~~-(l (l+1) (l (l+1) (l (l+1) (6 l(l+1)+113)-249)+214)+9) r^4 r_s\\
   &~~~~~~~~~~~~~-(l (l+1) (l (l+1) (l (l+1) (5 l (l+1)+21)-937)+655)+54) r^3 r_s^2\\
   &~~~~~~~~~~~~~+(l(l+1) (l (l+1) (9 l (l+1)+365)-2247)+351) r^2 r_s^3-(l (l+1) (10 l (l+1)+19)-972) r_s^5\\
   &~~~~~~~~~~~~~+(2268-l (l+1) (l (l+1) (5 l (l+1)-3)+817)) r r_s^4\Big] (\bm X^e)^2\\
\end{align*}

The terms with $p_2$ which are not already in $A_1$ or $A_2$ are in $A_3$:
\begin{align*}
    A_3=& \frac{8 r^5 \left(r^2-6 r r_s-3 r_s^2\right)}{4 l^2 (l+1)^2 r^2  \Lambda^2} (p_2')^2-\frac{4 r \sqrt{r r_s} \left(r_s+3 r\right)}{l (l+1) \Lambda} O_{q_1} p_2' -\frac{2 r \left(r^2-6 r r_s-3 r_s^2\right)}{l (l+1) \left(r_s+r\right)\Lambda} O_{p_1} p_2'\\
    &+ \frac{r \left(r^2-6 r r_s-3 r_s^2\right) \left((l-1) (l+2) \left(l^2+l-4\right) r^2-4 \left(l^2+l+1\right) r r_s-27 r_s^2\right)}{l^2 (l+1)^2 \left(r_s+r\right) \Lambda^3} p_2 p_2'\\
    &+\frac{2\sqrt{r r_s} \left(r^2-6 r r_s-3 r_s^2\right) \left(\left(l^2+l+6\right) r+3
   r_s\right)}{l^{3/2} (l+1)^{3/2} \left(r_s+r\right) \Lambda^2} \bm x^e{}' p_2'\\
   &+ \frac{\sqrt{r r_s}}{2 l^{3/2} (l+1)^{3/2} \left(r_s+r\right){}^2  r^2 \Lambda^3} \Big[3 \left(l (l+1) \left(l (l+1) \left(l^2+l+10\right)-1\right)+8\right) r^3 r_s^2\\
   &~~~~~~~~~~~~~+9 \left(l^2+l-38\right) r r_s^4-(l-1) l (l+1) (l+2) (5 l (l+1)+2) r^5\\
   &~~~~~~~~~~~~~+(l (l+1) (l (l+1) (2 l (l+1)+35)-96)+12) r^4 r_s\\
   &~~~~~~~~~~~~~+3 (l (l+1) (5 l (l+1)+2)-201) r^2 r_s^3-27 r_s^5\Big] \bm x^e p_2'\\
   &-\frac{\sqrt{r r_s} \left(r^2-6 r r_s-3 r_s^2\right) \left(\left(l^2+l+6\right) r+3
   r_s\right) \left((l-1) (l+2) \left(l^2+l-4\right) r^2-4 \left(l^2+l+1\right) r r_s-27
   r_s^2\right)}{2 l^{3/2} (l+1)^{3/2} \left(r_s+r\right){}^2 r^2 \Lambda^3} p_2 \bm x^e\\
   &+ \frac{1}{8 l^2 (l+1)^2 r
   \left(r_s+r\right){}^2 \Lambda^4} \Big[ \left(l^2+l-81\right) r_s^6\\
   &~~~~~~~~~~~~~+27\left(l (l+1) \left(l (l+1) \left(l (l+1)\left(l^2+l+27\right)+57\right)-1817\right)+2904\right) r^3r_s^3\\
   &~~~~~~~~~~~~~+27\left(l^2+l-31\right) \left(l^2+l+6\right) r r_s^5 +3 (l (l+1) (l (l+1)(3 l (l+1)+35)-803)+227) r^2 r_s^4\tag{\stepcounter{equation}\theequation}\\
   &~~~~~~~~~~~~~-(2 l (l+1)-7) \left(l (l+1) \left(l (l+1)
   \left(l^2+l-31\right)+98\right)-32\right) r^4 r_s^2\\
   &~~~~~~~~~~~~~-(l-1) (l+2) (l (l+1) (l (l+1) (7 l(l+1)-47)+154)-224) r^5 r_s\\
   &~~~~~~~~~~~~~-2 \left(l^2+l-2\right)^2 (3 l (l+1)-8) r^6\Big](p_2)^2\\
   &+ \frac{\sqrt{r r_s} \left(11 \left(l^2+l+16\right) r r_s^2-\left((l-1) (l+2) (4 l (l+1)-27)
   r^3\right)+(38 l (l+1)+17) r^2 r_s+45 r_s^3\right)}{2 l (l+1) r \left(r_s+r\right) \Lambda^2} O_{q_1}p_2\\
   &-\frac{\left(r^2-3 r_s \left(r_s+2 r\right)\right) \left((l-1) (l+2) \left(l^2+l-4\right) r^2-4\left(l^2+l+1\right) r r_s-27 r_s^2\right)}{2 l (l+1) \left(r_s+r\right){}^2 r \Lambda^2} O_{p_1} p_2\\
   &+\frac{2 \sqrt{r r_s}}{l^{3/2} (l+1)^{3/2}
   \left(r_s+r\right){}^3 16 r^4 \Lambda^4} \Big[-2 l (l+1) \left(l^2+l-2\right)^2 (l (l+1) (2 l (l+1)-11)+2) r^7\\
   &~~~~~~~~~~~~~+2 \left(l (l+1) \left(l (l+1) \left(l (l+1) \left(2 l (l+1)
   \left(l^2+l+7\right)-71\right)+43\right)+90\right)+72\right) r^5 r_s^2\\
   &~~~~~~~~~~~~~+(l-1) (l+2) (l (l+1) (l (l+1) (l (l+1) (4 l (l+1)+63)-224)+356)-48) r^6 r_s\\
   &~~~~~~~~~~~~~+3 (l (l+1) (l (l+1)
   (l (l+1) (5 l (l+1)-54)-536)+1978)-1748) r^4 r_s^3\\
   &~~~~~~~~~~~~~-6 (l (l+1) (l (l+1) (13 l (l+1)+179)-743)+162) r^3
   r_s^4-54 \left(l^2+l-173\right) r r_s^6\\
   &~~~~~~~~~~~~~-9 (2 l (l+1) (20 l (l+1)-83)-1937) r^2 r_s^5+729 r_s^7\Big] p_2 \bm x^e
\end{align*}

Finally, the remaining terms are in $A_4$:
\begin{align*}
   A_4=&-\frac{r_s \left(r^2-6 r r_s-3 r_s^2\right) \left(\left(l^2+l+6\right) r+3 r_s\right){}^2}{2 l (l+1)
   \left(r_s+r\right){}^2 r^2 \Lambda^2} (\bm x^e{}')^2 + \frac{2 r_s \left(\left(l^2+l+6\right) r+3 r_s\right)}{2\sqrt{l(l+1)} r \Lambda} O_{q_1}' \bm x^e\\
   &\frac{r_s}{4 l (l+1) r^4 \left(r_s+r\right){}^3 \Lambda^3} \Big[(l-1) l (l+1) (l+2) \left(l^2+l+6\right) (5 l (l+1)+2) r^6\\
   &~~~~~~~~~~~~~-2 \left(l (l+1) \left(l(l+1) \left(l (l+1) \left(l^2+l+16\right)+69\right)-276\right)+36\right) r^5 r_s\\
   &~~~~~~~~~~~~~-3 \left(l (l+1)
   \left(l (l+1) \left(l (l+1) \left(l^2+l+18\right)+94\right)-94\right)+60\right) r^4 r_s^2\\
   &~~~~~~~~~~~~~-27 \left(2 l (l+1) \left(l^2+l-5\right)-143\right) r^2 r_s^4-6 (l (l+1) (l (l+1) (4 l (l+1)+31)-96)-591) r^3r_s^3\\
   &~~~~~~~~~~~~~+1188 r r_s^5+81 r_s^6\Big] \bm x^e \bm x^e{}'\\
   &+\frac{2 r_s \left(r_s+3 r\right) \left(\left(l^2+l+6\right) r+3 r_s\right)}{\sqrt{l(l+1)}\left(r_s+r\right) r \Lambda}O_{q_1} \bm x^e{}'+\frac{\sqrt{r r_s} \left(r^2-6 r r_s-3 r_s^2\right) \left(\left(l^2+l+6\right) r+3 r_s\right)}{\sqrt{l}
   \sqrt{l+1} r^2 \left(r_s+r\right){}^2 \Lambda} O_{p_1}\bm x^e{}'\\
   &+ \frac{1}{8l (l+1) \sqrt{2(l+2)(l-1)} r^5 \left(r_s+r\right){}^4 \Lambda^2}\Big[2 l^2 (l+1)^2 \left(l^2+l-2\right)^3 r^7\\
   &~~~~~~~~~~~~~-(l-1) l (l+1) (l+2) (l (l+1) (l(l+1) (8 l (l+1)+79)-103)+78) r^6 r_s\\
   &~~~~~~~~~~~~~-(l (l+1) (l (l+1) (l (l+1) (2 l (l+1) (3 l(l+1)-5)-539)+1289)-204)+36) r^5 r_s^2\\
   &~~~~~~~~~~~~~+(l (l+1) (l (l+1) (l (l+1) (17 l (l+1)+435)+91)+279)-288) r^4
   r_s^3\\
   &~~~~~~~~~~~~~+3 (l (l+1) (l (l+1) (67 l (l+1)+115)-1971)+459) r^3 r_s^4\\
   &~~~~~~~~~~~~~+3 (l (l+1) (l (l+1) (4 l(l+1)+75)-745)+3960) r^2 r_s^5\\
   &~~~~~~~~~~~~~+9 (l (l+1) (8 l (l+1)+13)+693) r r_s^6+54 (2 l (l+1)+9) r_s^7\Big]\bm x^e \bm X^e\\
   &+ \frac{1}{4 \sqrt{l (l+1)}
   r^3 \left(r_s+r\right){}^2 \Lambda^2}\Big[-2 l (l+1) \left(l^2+l-2\right)^2 r^5\\
   &~~~~~~~~~~~~~+3 \left(l (l+1) \left(l (l+1)\left(l^2+l+9\right)-4\right)-24\right) r^3 r_s^2\\
   &~~~~~~~~~~~~~+15 \left(l^2+l-17\right) r r_s^4+(l-1) (l+2) (l (l+1) (5 l (l+1)+41)-6) r^4 r_s\tag{\stepcounter{equation}\theequation}\\
   &~~~~~~~~~~~~~+(l (l+1) (16 l (l+1)-17)-621) r^2 r_s^3-18 r_s^5\Big]O_{q_1} \bm x^e\\
   &+\frac{\sqrt{r r_s}}{4 \sqrt{l(l+1)}
   r^4 \left(r_s+r\right){}^3 \Lambda^2} \Big[-3 \left(l (l+1) \left(l (l+1) \left(l^2+l+10\right)-1\right)+8\right) r^3 r_s^2\\
   &~~~~~~~~~~~~~-9 \left(l^2+l-38\right) r r_s^4+(l-1) l (l+1) (l+2) (5 l (l+1)+2) r^5\\\
   &~~~~~~~~~~~~~-(l (l+1) (l (l+1) (2 l(l+1)+35)-96)+12) r^4 r_s\\
   &~~~~~~~~~~~~~-3 (l (l+1) (5 l (l+1)+2)-201) r^2 r_s^3+27 r_s^5\Big] O_{p_1} \bm x^e\\
   &-\frac{1}{32 l (l+1) r^6
   \left(r_s+r\right){}^4 \Lambda^4}\Big[l (l+1) \left(l^2+l-2\right)^4 \left(l^2+l+4\right) r^9\\
   &~~~~~~~~~~~~~-2 l (l+1) \left(l^2+l-2\right)^2 \left(l(l+1) \left(l^2+l-1\right) (5 l (l+1)+28)+68\right) r^8 r_s\\
   &~~~~~~~~~~~~~+\left(l (l+1) \left(l (l+1) \left(l (l+1)
   \left(21 l (l+1) \left(l^2+l+35\right)+88\right)-5296\right)+3808\right)-144\right) r^6 r_s^3\\
   &~~~~~~~~~~~~~+\left(l (l+1) \left(l (l+1) \left(l (l+1) \left(l^2+l+21\right)
   \left(l^2+l+58\right)-5420\right)-1466\right)+4824\right) r^4 r_s^5\\
   &~~~~~~~~~~~~~+27 \left(l (l+1) \left(2 l (l+1)
   \left(l^2+l+13\right)+11\right)+1429\right) r^2 r_s^7+81 \left(l^2+l+3\right) r_s^9\\
   &~~~~~~~~~~~~~-(l-1) l (l+1)
   (l+2) (l (l+1) (l (l+1) (l (l+1) (7 l (l+1)-23)-404)+1104)+56) r^7 r_s^2\\
   &~~~~~~~~~~~~~+(l (l+1) (l (l+1) (l (l+1) (l
   (l+1) (7 l (l+1)+488)+1122)-10559)+17708)-1440) r^5 r_s^4\\
   &~~~~~~~~~~~~~+3 (l (l+1) (l (l+1) (2 l (l+1) (2 l
   (l+1)+57)-45)-461)+20592) r^3 r_s^6\\
   &~~~~~~~~~~~~~+27 (l (l+1) (4 l (l+1)+25)+210) r r_s^8\Big](\bm x^e)^2
\end{align*}

\section{Perturbation Theory of Spherically Symmetric Spacetimes - Lagrangian Version}
\label{sec:PerturbLagrangian}

In this section we review the Lagrangian treatment of perturbation theory around a spherically symmetric background spacetime. This subject has been treated extensively in the literature from different points of view. In this section we follow closely the references \cite{3,5}.

The plan is the study of the linearized Einstein equations around a general spherically symmetric background. In the first section we derive the symmetry reduced vacuum Einstein equations. Then, we consider linear perturbations and expand the Einstein equations to first order. Finally we simplify the equations of motion for the perturbations and derive a master equation. In the case of a Schwarzschild background these equations have first been derived by Regge, Wheeler and Zerilli \cite{1,2}. In the approach presented here we derive a covariant version of the master equations. In this section we will use a different convention for the name of some of the indices which are not always compatible with the main text.

\subsection{Spherically Symmetric Degrees of Freedom}

We follow the formalism outlined in \cite{5} and consider the spacetime to be of product type $M_2 \times S^2$, with an arbitary 2-dimensional manifold $M_2$ and the 2-sphere $S^2$. On $M_2$ we define a metric $g$ and the associated Levi-Civita connection $\nabla$ which is torsion-free and compatible with $g$. For $S^2$ we have the usual metric $\Omega_{AB}$ on the 2-sphere and its associated Levi-Civita connection $D$. The indices on $M_2 \times S^2$ are Greek letters $\mu, \nu, \dots$, on $M_2$ small letters $a,b,\dots$ and for $S^2$ capital letters $A,B,\dots$. 

For the background degrees of freedom we study a symmetry reduced model compatible with spherical symmetry. Therefore, we take the ansatz
\begin{align}
	{}^4 g_{\mu \nu} \dd{x}^\mu \dd{x}^\nu &= g_{ab} \dd{x}^a \dd{x}^b + \gamma^2 \Omega_{AB} \dd{x}^A \dd{x}^B\,,
\end{align}
where $\gamma$ is an arbitrary, real-valued function on $M_2$. Inserting this into the Einstein equations we obtain the background equations of motion:
\begin{align}
    &2 \gamma \nabla_a \nabla_b \gamma + g_{ab}\qty(1 - \gamma_a \gamma^a - 2 \gamma \square \gamma ) =0\\
    &\gamma \square \gamma - \frac{1}{2}{}^2 R \gamma^2 = 0\,,
\end{align}
where we abbreviated $\gamma_a := \nabla_a \gamma$. The operator $\square = g^{ab} \nabla_a \nabla_b$ is the wave operator associated to the metric $g_{ab}$ and $R$ is the scalar curvature of $g_{ab}$.\\
We now study these equations more carefully. First, we define the quantity $f := \gamma_a \gamma^a$. Then, the first equation implies 
\begin{equation}
    \nabla_a \nabla_b \gamma = \frac{g_{ab}}{2 \gamma}(1-f)\,.
\end{equation}
The covariant derivative of the quantity $\gamma (1-f)$ vanishes because
\begin{align}
    \nabla_a (\gamma(1-f)) = \gamma_a (1-f) - 2 \gamma \gamma^b \nabla_a \gamma_b = 0\,.
\end{align}
Therefore, we have
\begin{align}
    f = 1 - \frac{r_s}{\gamma}\,,
\end{align}
where the integration constant turns out to be the Schwarzschild radius. This is seen when expanding $f$ in some explicit coordinate system for $M_2$ (see \cite{5}). It turns out that the function $f$ is precisely the function appearing in the Schwarzschild line element in standard Schwarzschild coordinates. 

\subsection{Perturbation Theory}

In the following we extend the previous discussion by including perturbations. We define them as

\begin{align}
    {}^4 g_{ab} &= g_{ab} + h_{ab} & {}^4 g_{aB} &= h_{aB} & {}^4 g_{AB} = \gamma^2 \Omega_{AB} + h_{AB}\,
\end{align}
where $h_{ab}$, $h_{aB}$ and $h_{AB}$ will be treated as linear perturbations. 

In order to formulate the Einstein equations, we need to expand the curvature tensors in terms of the perturbed metric. The perturbed Christoffel symbols to linear order are given by
\begin{align}
    \delta \Gamma^a_{bc} &= \frac{1}{2}\qty(\nabla_c h^a{}_b + \nabla_b h^a{}_c - \nabla^a h_{bc})\\
    \delta \Gamma^a_{bC} &= \frac{1}{2}\qty(D_C h^a{}_b + \nabla_b h^a{}_C - \nabla^ah_{bC}) - \frac{\gamma_b}{\gamma} h^a{}_C\\
    \delta \Gamma^a_{BC} &= \frac{1}{2}\qty(D_B h^a{}_C + D_C h^a{}_B - \nabla^a h_{BC}) + \gamma \gamma_m \Omega_{BC} h^{am}\\
    \delta \Gamma^A_{bc} &= \frac{1}{2\gamma^2}\qty(\nabla_b h^A{}_c + \nabla_c h^A{}_b - D^A h_{bc})\\
    \delta \Gamma^A_{bC} &= \frac{1}{2\gamma^2}\qty(D_C h^A{}_b - D^A h_{bC} + \nabla_b h^A{}_C ) - \frac{\gamma_b}{\gamma^3} h^A{}_C\\
    \delta \Gamma^A_{BC} &= \frac{1}{2\gamma^2}\qty(D_C h^A{}_B + D_B h^A{}_C - D^A h_{BC}) + \frac{\gamma_m}{\gamma} \Omega_{BC}h^{mA}
\end{align}
We use the convention that the indeces $a,b,\dots$ are moved with the metric $g_{ab}$ and the indices $A,B,\dots$ with the metric $\Omega_{AB}$. Using the perturbed Christoffel symbols, the perturbed Ricci tensor is given by

\begin{align}
\begin{split}
    \delta R_{ab} &= \nabla_{m} \qty(\nabla_a h^m{}_b + \nabla_b h^m{}_a - \nabla^m h_{ab})+\frac{\gamma_m}{\gamma} \qty(\nabla_a h^m{}_b + \nabla_b h^m{}_a - \nabla^m h_{ab})-\frac{1}{2} \nabla_{a} \nabla_{b} h^m{}_m\\
    &- \frac{1}{2 \gamma^2} D^{M} D_{M} h_{a b}+\frac{1}{2 \gamma^2} D_{M}\left(\nabla_{a} h^{M}{}_{b}+\nabla_{b} h^{M}_{a}\right)-\frac{1}{2 \gamma^{2}} \nabla_{a} \nabla_{b} h^{M}{}_{M}\\
    &+\frac{1}{2 \gamma^{3}}\left(\gamma_{a} \nabla_{b} h^M{}_{M}+\gamma_{b} \nabla_{a} h^M{}_{M}\right)-\frac{1}{\gamma^{4}}\left(\gamma_{a} \gamma_{b}-\gamma \nabla_{a} \nabla_{b} \gamma\right) h^M{}_{M}
    \end{split}\\
    \begin{split}
    \delta R_{aB} &= \frac{1}{2} D_{B}\left(\nabla_{m} h^m{}_{a}-\nabla_{a} h^m{}_{m}+\frac{1}{\gamma} \gamma_{a} h^m{}_{m}\right)-\frac{1}{2}\left(\square h_{a B}-\nabla_{m} \nabla_{a} h^m{}_{B}\right)\\
    &-\frac{1}{\gamma}\left(\gamma_{a} \nabla_{m} h^m{}_{B}-\gamma_{m} \nabla_{a} h^m{}_{B}\right)-\frac{1}{\gamma^{2}}\left(\gamma_{a} \gamma_{m}+\gamma \nabla_{a} \nabla_{m} \gamma\right) h^m{}_{B}+\frac{1}{2 \gamma^{2}} D^{M}\left(D_{B} h_{a M}-D_{M} h_{a B}\right)\\
    &+\frac{1}{2 \gamma^{2}} \nabla_{a}\left(D_{M} h^M{}_{B}-D_{B} h^M{}_{M}\right)-\frac{1}{\gamma^{3}} \gamma_{a}\left(D_{M} h^M{}_{B}-D_{B} h^M{}_{M}\right)\end{split}\\
    \delta R_{AB} &= \Omega_{A B}\left[\gamma \gamma_{m} \nabla_{n}\left(h^{m n}-\frac{1}{2} g^{m n} h^k{}_{k}\right)+\left(\gamma_{m} \gamma_{n}+\gamma \nabla_{m} \nabla_{n} \gamma\right) h^{m n}\right]-\frac{1}{2} D_{A} D_{B} h^m{}_{m}\nonumber\\
    &+\frac{1}{2} \nabla_{m}\left(D_{A} h^m{}_{B}+D_{B} h^m{}_{A}\right) +\frac{1}{\gamma} \gamma_{m} \Omega_{A B} D_{M} h^{m M}+\frac{1}{2\gamma^{2}} D_{M}(D_A h^M{}_B + D_B h^M{}_A- D^M h_{AB})\\
    &-\frac{1}{2} \square h_{A B} -\frac{1}{2 \gamma^{2}} D_{A} D_{B} h^M{}_{M} + \frac{1}{\gamma} \gamma^{m} \nabla_{m}\left(h_{A B}-\frac{1}{2} \Omega_{A B} h^M{}_{M}\right) -\frac{2}{\gamma^{2}} \gamma^{m} \gamma_{m}\left(h_{A B}-\frac{1}{2} \Omega_{A B} h^M{}_{M}\right)\nonumber
\end{align}

The next computations are most easily done in terms of tensor harmonics. For a review of spherical harmonics see appendix \ref{sec:sphericalHarmonics}.

\subsubsection{Gauge Invariant Variables in Terms of Tensor Harmonics}

The perturbed metric is expanded into the tensor harmonics as follows
\begin{align}
    h_{ab} &= \sum_{l,m} h^{lm}_{ab} L^{lm}\\
    h_{aB} &= \sum_{l,m} \sum_{I \in \{e,o\}} h_a^{I,lm} (L^{lm}_I)_{B}\\
    h_{AB} &= \gamma^2 \sum_{lm}\qty( h^{h,lm} L^{lm} \Omega_{AB} +\sum_{I \in \{e,o\}} h^{I,lm} (L^{lm}_{I})_{AB})
\end{align}
Here, $o$ stands for the odd, $e$ for the even and $h$ for the trace part of the respective component. In this section we only consider the modes with $l\geq 2$. They are relevant for the derivation of the Regge-Wheeler and Zerilli equations. The modes with $l=1$ are discussed in the literature (see \cite{5}) and we will not repeat their analysis here.

The Einstein equations simplify dramatically if presented in terms of gauge invariant variables. 
For that consider gauge transformations (i.e. spacetime diffeomorphisms) 
generated by the vector field $\xi = (\xi_a L, \xi_A =\xi^o L^o_A + \xi^e L^e_A)$. The metric perturbations transform according to
\begin{align}
    h_{ab} &\to h_{ab} - \nabla_a \xi_b - \nabla_b \xi_a\\
    h_{aB} &\to h_{aB} - \nabla_a \xi_B - D_B \xi_a + \frac{2 \gamma_a}{\gamma}\xi_B\\
    h_{AB} &\to h_{AB} - D_A \xi_B - D_B \xi_A - 2 \gamma \gamma^a \xi_a \Omega_{AB}
\end{align}
In the following we suppress the labels of the spherical harmonics. The equations are valid for any allowed value of $l$ and $m$. The expansion coefficients of the perturbed metric transform under gauge transforamtion by the vector field $\xi$ as
\begin{align}
    h_{ab} &\to h_{ab} - \nabla_a \xi_b - \nabla_b \xi_a\\
    h^e_a &\to h^e_a - \sqrt{l(l+1)}\xi_a - \nabla_a \xi^e + \frac{2}{\gamma} \gamma_a \xi^e\\
    h^o_a &\to h^o_a - \nabla_a \xi^o + \frac{2}{\gamma}\gamma_a \xi^o\\
    h^{h} &\to h^{h} + \frac{l(l+1)}{\gamma^2}\xi^e - \frac{2}{\gamma} \gamma^a \xi_a\\
    h^e &\to h^e - \frac{2}{\gamma^2}\sqrt{\frac{(l+2)(l-1)}{2}}\xi^e\\
    h^o &\to h^o - \frac{2}{\gamma^2}\sqrt{\frac{(l+2)(l-1)}{2}}\xi^o
\end{align}

For the construction of the gauge invariant variables, we form combinations of the perturbations which transform trivially under gauge transformations. The following quantities are gauge invariant:
\begin{align}
    k_{ab} &= h_{ab} - \frac{1}{\sqrt{l(l+1)}}\nabla_a \epsilon_b - \frac{1}{\sqrt{l(l+1)}}\nabla_b \epsilon_a\\
    \tilde h_a &= h^o_a - \sqrt{\frac{2}{(l+2)(l-1)}}\frac{\gamma^2}{2}\nabla_a h^o\\
    K &= h^{h} + \frac{1}{2}l(l+1) h^e - \frac{2}{\sqrt{l(l+1)}\gamma} \gamma^a \epsilon_a\,.
\end{align}
We introduced the variable $\epsilon_a$ defined as
\begin{equation}
\epsilon_a := h^e_a - \frac{1}{2} \sqrt{\frac{2}{(l+2)(l-1)}}\gamma^2 \nabla_a h^e\,. 
\end{equation}
It transforms under gauge transformations as $\epsilon_a \to \epsilon_a - \sqrt{l(l+1)}\xi^e_a$.

\subsubsection{Odd Parity Master Equation}
The Einstein equations for the odd parity sector in gauge invariant variables are given by
\begin{align}
    &\nabla_m \tilde h^m = 0\,,\\
    &\frac{1}{2}\qty(\nabla_m \nabla_a \tilde h^m - \square \tilde h_a) - \frac{1}{\gamma}\qty(\gamma_a \nabla_m - \gamma_m \nabla_a) \tilde h^m - \frac{1}{\gamma^2}\qty(\gamma_a \gamma_m + \gamma \nabla_a \nabla_m \gamma) \tilde h^m\\
    &+ \frac{(l+2)(l-1)}{2 \gamma^2} \tilde h_a + \frac{\square \gamma}{\gamma} \tilde h_a + \frac{\gamma^m \gamma_m}{\gamma^2} \tilde h^a = 0\,.
\end{align}
As noted by \cite{3}, the second equation can be put into the simpler form
\begin{align}
    \nabla^c(\gamma^4 \nabla_{[a}( \gamma^{-2} \tilde h_{c]})) + \frac{1}{2} (l+2)(l-1) \tilde h_a = 0\,,\\
    \frac{1}{2} \epsilon_{ac} \nabla^c(\gamma^4 \epsilon^{de} \nabla_d (\gamma^{-2} \tilde h_e)) + \frac{1}{2} (l+2)(l-1) \tilde h_a = 0\,.
\end{align}

From here we can derive the master equation for the odd parity perturbations. The master variable is defined as  
\begin{equation}
    \psi = \gamma^3 \epsilon^{ab} \nabla_a\qty(\gamma^{-2} \tilde h_b).
\end{equation}
Plugging this into the equation and applying the operator $\epsilon^{ba} \nabla_b \qty(\gamma^{-2}\cdot)$ gives the odd parity master equation:
\begin{equation}
   - \frac{1}{2}\gamma \nabla_a\qty(\gamma^{-2} \nabla^a\qty(\gamma \psi)) + \frac{(l+2)(l-1)}{2\gamma^2} \psi = 0.
\end{equation}
After expanding the derivatives, we obtain
\begin{equation}
    \square \psi + \qty(-\frac{l(l+1)}{\gamma^2} + \frac{1}{\gamma^2}\qty(2 - 2\gamma_a \gamma^a + \gamma \square \gamma)) \psi = 0\,.
\end{equation}

This is the Regge-Wheeler equation for the odd parity perturbations. Using the solution found for the background degrees of freedom we obtain
\begin{align}
    \square \psi - V_{\mathrm{RW}} \psi = 0\,,
\end{align}
with the Regge-Wheller potential
\begin{align}
	V_{\mathrm{RW}} = \frac{l(l+1)\gamma - 3 r_s}{\gamma^3}\,.
\end{align}

\subsubsection{Even Parity Master Equation}

We now study the even parity sector. The linearized Einstein equations are given by
\begin{align}
    \begin{split}
    &\frac{1}{2}\qty(\nabla_m \nabla_a k^m_b + \nabla_m \nabla_b k^m_a - \square k_{ab}) + \frac{\gamma_m}{\gamma}\qty(\nabla_a k^m_b + \nabla_b k^m_a - \nabla^m k_{ab}) - \frac{1}{2}\nabla_a \nabla_b k^m_m\\
    &+ \frac{l(l+1)}{2\gamma^2} k_{ab} - \nabla_a \nabla_b K  - \frac{1}{\gamma}\qty(\gamma_a \nabla_b + \gamma_b \nabla_a)K+ k_{ab}\qty(- \frac{ {}^2 R }{2} -\frac{1}{\gamma^2} + \frac{\gamma^m\gamma_m}{\gamma^2} + \frac{2 \square \gamma}{\gamma}) \\
    &- \frac{1}{2}g_{ab} \Big(\nabla_m \nabla_n k^{mn} - \square k^m_m + 2 \frac{\gamma_m}{\gamma}\qty(2 \nabla_n k^{mn} - \nabla^m k^n_n) + \frac{2}{\gamma^2}\qty(\gamma_m \gamma_n + 2 \gamma \nabla_m \nabla_n \gamma) k^{mn}+ \frac{l (l + 1)}{\gamma^2} k^m_m\\
    &- 2 \square K - 6\frac{\gamma^m}{\gamma} \nabla_m K + \frac{(l+2)(l-1)}{\gamma^2}K - {}^2 R_{mn} k^{mn}\Big) = 0\,,
    \end{split}\\
    &\nabla_m  k^m_a - \nabla_a k^m_m + \frac{\gamma_a}{\gamma} k^m_m - \nabla_a  K = 0\,,\\
    \begin{split}
    &  \frac{\gamma}{2}\qty(\nabla^a(\gamma \nabla_a k^m_m)  - \gamma \nabla_a \nabla_b  k^{ab} - 2 \gamma_a \nabla_b  k^{ab} - 2\nabla_a \nabla_b \gamma  k^{ab})\\
    &- \frac{l(l+1)}{4}  k^m_m + {}^2 R \frac{ \gamma^2}{4}  k^m_m + \frac{\gamma^2}{2}\square  K + \gamma \gamma^a\nabla_a  K = 0\,,
    \end{split}\\
    &  k^a_a = 0\,.
\end{align}

An interesting identity can be derived by noting that the Einstein tensor on any two dimensional manifold vanishes. For this consider a perturbed metric $g_{ab} + h_{ab}$ on $M_2$. Then the Ricci curvature to first order in the perturbations $h_{ab}$ is given by
\begin{equation}
    {}^{(2)}R_{ab} = \frac{1}{2} g_{ab} {}^{(2)}R(g) + \frac{1}{2}(2 \nabla^c \nabla_{(a} h_{b)c} - \square h_{ab} - \nabla_a \nabla_b h^c_c)\,,
\end{equation}
where ${}^{(2)}R(g)$ is the unperturbed Ricci scalar of $g$. The vanishing Einstein tensor reads
\begin{equation}
    2 {}^{(2)}G_{ab} = \nabla^c \nabla_a h_{bc} + \nabla^c \nabla_b h_{ac} - \square h_{ab} - \nabla_a \nabla_b h^c_c + g_{ab} (\square h^c_c - \nabla^c \nabla^d h_{cd}) + \frac{1}{2} {}^{(2)}R(g_{ab} h^c_c - 2 h_{ab}) = 0\,.
\end{equation}
We use this identity in the first equation. It removes all the terms with second derivatives of $k_{ab}$. In addition, the last equation forces the trace of $k_{ab}$ to vanish. Therefore, we obtain the following system of equations
\begin{align}
    &\frac{\gamma^m}{\gamma}\qty(\nabla_a k_{mb} + \nabla_b k_{am} - \nabla_m k_{ab})+ \frac{l(l+1)}{2\gamma^2} k_{ab} - \nabla_a \nabla_b K  + k_{ab}\qty( -\frac{1}{\gamma^2} + \frac{\gamma^m\gamma_m}{\gamma^2} + \frac{2 \square \gamma}{\gamma}) - \frac{1}{\gamma}\qty(\gamma_a \nabla_b + \gamma_b \nabla_a)K\nonumber\\
    &- \frac{1}{2}g_{ab} \Big(4 \frac{\gamma_m}{\gamma} \nabla_n k^{mn} + \frac{2}{\gamma^2}\qty(\gamma_a \gamma_b + 2 \gamma \nabla_a \nabla_b \gamma) k^{ab} - 2 \square K - 6\frac{\gamma_a}{\gamma} \nabla^a K + \frac{(l+2)(l-1)}{\gamma^2}K\Big) = 0\,,\\
    &\nabla_m  k^m_a - \nabla_a  K = 0\,,\\
    &- \frac{\gamma}{2}\qty(\gamma \nabla_a \nabla_b  k^{ab} + 2 \gamma_a \nabla_b  k^{ab} + 2\nabla_a \nabla_b \gamma  k^{ab})+ \frac{\gamma^2}{2}\square  K + \gamma \gamma^a\nabla_a  K = 0\,.
\end{align}

The second equation can be used to relate $\nabla_a K$ and $\nabla_m k^m_a$. Plugging this relation into the third equation we obtain
\begin{equation}
    \nabla_a \nabla_b \gamma k^{ab} = 0\,.
\end{equation}
Note that this equation follows from what we observed for the background variables. We showed that $\nabla_a \nabla_b \gamma$ is proportional to $g_{ab}$. 

The derivation of the master equation is more involved compared to the odd parity sector. We first derive three preliminary equations which are then combined to the master equation. We start by taking the trace of the first equation 
\begin{equation}
    \square K - \frac{\lambda}{\gamma^2} K - \frac{2}{\gamma^3}\gamma^a Z_a = 0\,,
    \label{eq:S1}
\end{equation}
where we defined $Z_a := \gamma (\gamma^b k_{ab} - \gamma \nabla_a K)$ and $\lambda = (l+2)(l-1)$.

The second equation is obtained by contracting the first equation with $\gamma^a$. Then use the equation \eqref{eq:S1} to remove the term proportional to $\square K$. We have
\begin{equation}
    \nabla_b (\gamma^a Z_a) + \frac{l(l+1)}{2\gamma} Z_b + \frac{\gamma}{2} 3(1 - \gamma_a \gamma^a)\nabla_b K + \frac{1}{2}\lambda \nabla_b (\gamma K) = 0\,.
    \label{eq:S2}
\end{equation}

The last equation is a direct consequence of the original second equation contracted with $\gamma^a$ and reads
\begin{equation}
    \nabla^a Z_a - \frac{\gamma^a}{\gamma} Z_a + \gamma^2 \square K = 0\,,
    \label{eq:S3}
\end{equation}

The even parity master equation is now derived by applying the operator $\nabla^b$ on Equation \eqref{eq:S2}. Together with the equation \eqref{eq:S1} to remove $\square K$ and equation \eqref{eq:S3} to remove $\nabla^a Z_a$ we obtain the following equation
\begin{equation}
    \square(2 \gamma^a Z_a + \lambda \gamma K) + \qty[- \frac{l(l+1)}{\gamma^2} + \frac{3}{\gamma^2}(1 - \gamma_a \gamma^a)]\qty(2 \gamma^a Z_a + \lambda \gamma K ) = 0\,.
\end{equation}

Therefore we have a master equation for the master variable $\psi = 2 \gamma^a Z_a + \lambda \gamma K$. The potential is the same as the one in the odd parity case. In the literature usually a different form of the master equation is discussed. It uses the covariant form of the Zerilli variable defined as
\begin{equation}
    \psi_Z = \frac{1}{l(l+1)}\qty(\gamma K + \frac{1}{\Lambda} \gamma^a Z_a)\,,
\end{equation}
where $\Lambda = \frac{1}{2}(l+2)(l-1) + \frac{3 r_s}{2\gamma}$. The Zerilli variable satisfies a wave equation with a different potential. We have
\begin{equation}
    \square \psi_Z  - V_{\mathrm{Z}}\psi_Z = 0\,,
\end{equation}
with the Zerilli potential 
\begin{equation}
    V_{\mathrm{Z}}= \frac{1}{2\Lambda^2 r^2}\qty(\lambda^2(\lambda+2) + 3 \lambda^2 \frac{r_s}{r} + 9 \lambda \frac{r_s^2}{r^2} + 9 \frac{r_s^3}{r^3})\,.
\end{equation}
The two master variables are related through a simple transformation, see \cite{Chandrasekhar1983}. We have
\begin{equation}
    \psi = - 6 r_s \gamma^a \nabla_a \psi_Z + \qty((l+2)(l+1)l(l-1) + \frac{9 r_s^2}{\gamma^2 \Lambda} \gamma^a \gamma_a) \psi_Z\,.
\end{equation}

This completes the treatment of linear perturbation theory around a spherical symmetric background spacetime. 
We reduced the problem to two wave equations for two master variables
\begin{align}
    \square \psi - V \psi = 0\,.
\end{align}
The potentials $V$ are the Regge-Wheeler potential $V_{\mathrm{RW}}$ for the odd parity case and the Zerilli potential $V_{\mathrm{Z}}$ for the even parity perturbations.


\end{document}